\documentclass[letterpaper,12pt]{article}

\usepackage{natbib}
\usepackage{amsmath,amsthm,amsfonts}
\usepackage{enumitem}

\usepackage{algpseudocode}
\usepackage{algorithm}
\usepackage{graphicx}
\usepackage{xcolor}

\usepackage{tikz}
\usepackage{tikz-cd}

\usepackage{xargs}
\usepackage{booktabs}
\usepackage{multirow}
\usepackage[labelformat=simple]{subcaption}

\usepackage{changes}

\usepackage{aliascnt}

\newtheorem{theorem}{Theorem}[section]

\newaliascnt{proposition}{theorem}
\newtheorem{proposition}[proposition]{Proposition}
\aliascntresetthe{proposition}

\newaliascnt{corollary}{theorem}
\newtheorem{corollary}[corollary]{Corollary}
\aliascntresetthe{corollary}

\newaliascnt{lemma}{theorem}
\newtheorem{lemma}[lemma]{Lemma}
\aliascntresetthe{lemma}

\theoremstyle{definition}
\newaliascnt{definition}{theorem}
\newtheorem{definition}[definition]{Definition}
\aliascntresetthe{definition}

\theoremstyle{remark}
\newtheorem*{remark}{Remark}

\usepackage[
    colorlinks=true,
    citecolor=blue,
    linkcolor=red
]{hyperref}

\usepackage[noabbrev]{cleveref}

\crefname{theorem}{theorem}{theorems}
\Crefname{theorem}{Theorem}{Theorems}

\crefname{proposition}{proposition}{propositions}
\Crefname{proposition}{Proposition}{Propositions}

\crefname{corollary}{corollary}{corollaries}
\Crefname{corollary}{Corollary}{Corollaries}

\crefname{lemma}{lemma}{lemmas}
\Crefname{lemma}{Lemma}{Lemmas}

\crefname{definition}{definition}{definitions}
\Crefname{definition}{Definition}{Definitions}

\setlist{
    itemsep=0ex,
    topsep=1ex,
    parsep=0ex
}

\setlist[enumerate]{
    label=(\arabic*)
}

\def\Grp#1{\left(#1\right)}
\def\Cbr#1{\left\{#1\right\}}

\def\Flr#1{\left\lfloor#1\right\rfloor}
\def\Cil#1{\left\lceil#1\right\rceil}

\def\Sb#1{\sb{(#1)}}
\def\Sp#1{\sp{(#1)}}

\def\Iff{\Leftrightarrow}   

\def\ncol{\operatorname{ncol}}
\def\nrow{\operatorname{nrow}}
\def\ker{\operatorname{ker}}
\def\image{\operatorname{im}}
\def\pred{\operatorname{pred}}

\def\sppt{\operatorname{sppt}}
\def\succ{\operatorname{succ}}
\def\Id{\operatorname{Id}}

\makeatletter\def\@cleandot{\@ifnextchar.{}{\@ifnextchar,{.}{\@ifnextchar;{.}{\@ifnextchar?{.}{\@ifnextchar:{.}{\@ifnextchar!{.}{\@ifnextchar'{.}{\@ifnextchar){.}{\@ifnextchar({.}{\@ifnextchar-{.}{\@ifnextchar\/{.}{\@ifnextchar\\{.}{.\ }}}}}}}}}}}}}
\def\lhs{l.h.s\@cleandot}
\def\rhs{r.h.s\@cleandot}
\makeatother

\def\Ints{\mathbb{Z}}
\def\Reals{\mathbb{R}}

\def\Cal#1{\mathcal{#1}}

\newcommandx*\cprod[3][1=1]{#2_{#1}\cdots#2_{#3}}
\newcommandx*\cset[3][1=1]{#2_{#1} \ldots #2_{#3}}
\newcommandx*\cum[3][1=1]{#2_{#1}+\cdots+#2_{#3}}
\newcommandx*\eno[3][1=1]{#2_{#1},\ldots,#2_{#3}}
\newcommandx*\seqop[4][1=1]{#2_{#1}#3\cdots#3 #2_{#4}}

\title{\Large\bf Recursive Computation of Path Homology for Stratified
  Digraphs}
\author{Zhengtong Zhu and Zhiyi Chi\footnote{Department of Statistics,
    University of Connecticut, Storrs, CT 06250.}\ \footnote{Email:
    zhengtong.zhu, zhiyi.chi@uconn.edu}
}
\date{August 4, 2026}

\begin{document}
\maketitle
\begin{abstract}
  Stratified digraphs are popular models for feedforward neural
  networks.  However, computation of their path homologies has been
  limited to low dimensional ones due to high computational
  complexity.  A recursive algorithm is proposed to compute certain
  high-dimensional (reduced) path homologies of stratified digraphs.
  By recursion on matrix representations of homologies of subgraphs,
  the algorithm efficiently computes the full-depth path homology of a
  stratified digraph, i.e. homology with dimension equal to the depth
  of the graph.  The algorithm can be used to compute the maximal path
  homology of acyclic digraphs, i.e., path homology with dimension
  equal to the maximum path length of a graph.  Numerical experiments
  show that the algorithm has a significant advantage over the general
  algorithm in computation time as the depth of stratified digraph
  increases.
\end{abstract}

\section{Introduction} \label{s:intro}
Directed graphs, or digraphs, are regularly used in applications.  To
grapple with the complexity of large digraphs, the analysis of their
homologies has gained traction in recent years \citep[cf.][and
references therein]{ballester2023topological}.  This represents a
new development of topological data analysis (TDA), an area that has
been intensively studied since its introduction by \cite
{edelsbrunner2002topological}.  In TDA, the most common approach is to treat data as a
cloud of points in a metric space and calculates homology groups of an
associated simplicial complex, typically a \v{C}ech or Vietoris--Rips complex
\citep {li2020finding, berry2020functional, biscio2019accumulated,
  moon2023hypothesis, rathore2021topoact, chazal2014persistence,
  chazal2014convergence, edelsbrunner2022computational,
  wasserman2018topological, loiseaux2024framework}.  Open source software to compute the homologies of these complexes has become readily available \citep{chomp,gudhi:urm}.  The approach has
been extended to neural networks by defining simplices as cliques of
neurons that are in proximity under certain (pseudo-)metric \citep
{watanabe2022topological, masulli2016topology}.  In addition,
Dowker complexes can be defined based on pairwise relations between
neurons that are specified according to the weights of their
connections \citep{dowker1952homology, chowdhury2018functorial}.  In
these approaches, as the underlying simplices are undirected, the ensuing calculation
of homologies has nothing to do with edge orientations.  To a large
degree, the homologies can be characterized as simplicial homologies
for undirected graphs \citep {chen2001graph,
  ivashchenko1994contractible}.  To better capture the inherent
asymmetry of digraphs due to edge orientations, it is desirable to
incorporate edge orientations in an essential manner.  One way is to
treat a digraph as a directed one-dimensional complex
\citep{rieck2018neural}.  However, this leads to trivial homologies of
all dimensions higher than 1.  Another way is to consider directed
flag complexes, which have directed cliques as simplices
\citep{reimann2017cliques, masulli2016topology}.  The resulting
homologies have more structure, however, as shown by \cite
{chowdhury2019path}, all two- and higher-dimensional ones can be
trivial even for digraphs with high complexity.  Partly aimed to
address these limitations, path homology was introduced by \cite
{grigoryan2012homologies} and has revealed numerous nontrivial
properties \citep{grigoryan2017homologies}.

The focus of the paper is numerical computation of path homologies for
deep feedforward networks, i.e., artificial neural networks that
process input data in multiple stages with each stage being done by a
completely separate set, or ``layer'' of neurons solely based on the
output from the previous stage \citep{murphy:22:mit}.  Theoretically,
\cite{chowdhury2019path} showed that the path homologies of a
feedforward network can be calculated in closed form if it is fully
connected, i.e., every pair of its neurons in adjacent layers are
connected.  Numerically, since feedforward networks can be
represented as stratified digraphs \cite[also see \Cref {d:stratified}
below] {rieck2018neural}, their path homologies can be computed using
an algorithm designed for digraphs in general \citep
{chowdhury2017persistent}.  This is the approach taken by most of the
existing methods.  However, the general algorithm does not fully
exploit the structural characteristics of a stratified digraph.  Since
the algorithm computes the $p$-dimensional path homology of a digraph
of $n$ vertices with complexity $O(n^{6+3p})$, it may have difficulty
computing high-dimensional homologies.  For one-dimensional
homologies, the algorithm proposed by \cite {dey2022efficient}
leverages arboricity and 1-boundary group structures to achieve
efficient computation, especially for digraphs that are dense or have
low arboricity.  However, it is unclear if the algorithm can be
generalized to high-dimensional homologies.  For deep feedforward
networks, one can also compute homologies that are different from path
homologies.  Due to the high computational complexity, existing methods
are restricted to three- or lower-dimensional homologies \citep
{rieck2018neural, perez2021characterizing}. 

For stratified digraphs, this paper considers certain high-dimensional
(reduced) path homologies instead of low-dimensional ones.  First, we
recall the definition of a stratified digraph and introduce some
related terms.
\begin{definition}  \label{d:stratified}
  Let $G=(V,E)$ be a digraph.  If $V$ can be partitioned into
  nonempty subsets $\eno[0] KL$, such that $E\subset\{(u,v): u\in
  K_i, v\in K_{i+1}, 0\le i<L\}$, then $G$ is said to be
  \emph{stratified\/}, with the $K_i$'s being its \emph{layers\/} and
  $L$ its \emph{depth\/}.
\end{definition}
\begin{remark}
  1) For convenience, we define layers differently from \cite
  {rieck2018neural}.  In the latter, the $i$th layer refers to the
  subgraph of $G$ consisting of $K_i$, $K_{i+1}$, and $E_i:=\{(u,v)
  \in E: u\in K_i, v\in K_{i+1}\}$.
  
  2) In general, layers are not uniquely defined.  For example, if a
  node has no incoming or outgoing edge, then it can be put into any
  layer,  Also, if $E_j=\emptyset$ for some $0\le j<L$, then $K'_0 =
  K_{j+1}$, \ldots, $K'_{L-j-1} = K_L$, $K'_{L-j} = K_0, \ldots$,
  $K'_L = K_j$ can be regarded as the $0$th, $1$st, \ldots, and $L$th 
  layers of $G$, respectively.  However, for a feedforward network,
  the layers can be uniquely specified by its architecture.
\end{remark}

Let $G$ be a stratified digraph of depth $L$ as in \Cref
{d:stratified}.  This paper proposes a recursive algorithm to compute
its $L$-dimensional, or \emph{full-depth} path homology.  The
algorithm is based on two facts.  First, as $G$ has a trivial space of
$L$-boundaries, its full-depth path homology is the same as its
full-depth cycle space, i.e., the space of $L$-cycles.  Second, the
full-depth cycle space is Markovian in the sense that full-depth
cycles of $G$ can be expressed linearly in terms of those of the
subgraph between the top and penultimate layers, i.e., $(\cup_{i<L}
K_i, \cup_{i<L-1} E_i)$, with the coefficients determined by the
subgraph ``at the bottom'', i.e., $(K_{L-1}\cup K_L, E_{L-1})$.  This
allows the full-depth cycle space to be updated layer by layer without
tracking actual paths of cycles, hence significantly reducing
computation.  On the other hand, the algorithm allows one to opt for
tracking the actual paths of a basis of the full-depth homology.  The
computation time remains less than that of the general algorithm for
deep stratified digraph, although at the same order of magnitude.

In some cases, $G$ only has paths shorter than its depth, so its
full-depth path homology is trivial.  One may then wish to find its
$\ell(G)$-dimensional, or \emph{maximal\/} path homology instead,
where $\ell(G)$ denotes the length of the longest directed path in
$G$.  Indeed, any directed acyclic graph (DAG) has a well defined
maximal path homology.  It will be seen that the calculation of
maximal path homology can be reduced to that of full-depth path
homology.

The rest of the paper is organized as follows.  \Cref{s:preliminaries}
provides a brief background on path homology and describes the general
algorithm of \cite{chowdhury2021path} to compute the path homologies
of a digraph.  \Cref{s:basic} establishes preliminary results on path
homologies of digraphs.  \Cref{s:recursive} proves the main results
that lead to the recursive computation of full-depth path homologies
of stratified digraphs.  \Cref{s:algorithms} presents an algorithm for
the computation as well as several auxiliary algorithms.  \Cref
{s:experiments} reports some numerical experiments on the recursive
algorithm.  In the first set of experiments, the algorithm was
compared with an implementation of the general algorithm furnished by
\cite{pyproject2022}.  Both were applied to compute the full-depth
homologies of feedforward networks randomly generated with depth up to
10 and number of neurons in each layer up to 10.  The run
times and memory usages of the algorithms for the tasks are compared.
In the second set of experiments, to see if the recursive algorithm
had any potential application to persistent homology, it was applied
to compute Betti numbers of full-depth path homologies of nested
stratified digraphs.  The plot of Betti number \emph{vs.\/} the
parameter of the nested graphs, or ``Betti curve'', is a useful
graphical tool in TDA of digraphs \citep {chowdhury2019path}.  Note,
however, that persistent homology of digraphs is concerned with many
other issues than just Betti numbers \citep{masulli2016topology,
  watanabe2022topological, chowdhury2018functorial}, which our
algorithm is  unable to address.  All the code developed in the paper
is open-source; see \cref{s:experiments} for detail. 
\Cref{s:conclusion} concludes with a brief discussion.

\section{Preliminaries}\label{s:preliminaries}
\subsection{Path homology}
There are several works in the literature that cover path
homologies in detail \citep{grigoryan2012homologies,
  grigoryan2017homologies, chowdhury2021path}.  In this
paper, we only consider reduced path homologies with real
coefficients.

Recall that the vector space over $\Reals$ generated by a finite set
$S$ is $\Reals^S = \{(v_s)_{s\in S}: v_s\in\Reals\}$.  Each $\eta\in
\Reals^S$ can be uniquely written as $\sum_{s\in S} a_s e_s$, where
$a_s\in\Reals$ and $e_s$ is the vector $(v_t)_{t\in S}$ with $v_t = 0$
for all $t\ne s$ and $v_s=1$.  By convention,
$\Reals^\emptyset=\{0\}$.

Let $G=(V,E)$ be a digraph, where $V$ is a non-empty finite set.
\begin{definition}[Paths on vertices]
  For integer $p\ge0$, an \emph{elementary $p$-path\/} on $V$ is any
  sequence $(\eno[0] x p)$ of $p+1$ vertices of $V$; $p$ is referred
  to as its \emph{length\/}, and $x_i$ the element at its $i$th
  position.  Let $\Lambda_p = \Lambda_p(V)$ be the vector space
  over $\Reals$ generated by the set of elementary $p$-paths on $V$.
  Each element of $\Lambda_p$ is called a \emph{$p$-path} and can be
  uniquely written as $\sum_s a_s e_s$, where $a_s\in\Reals$ and the
  sum is over all elementary $p$-paths $s$.  We will formally write
  the $p$-path as $\sum_s a_s s$ instead.  Since we only consider
  reduced homologies, define $\Lambda_{-1} \equiv\Reals$ and for
  $p<-1$, define $\Lambda_p\equiv\{0\}$.
\end{definition}
By the definition, the set of elementary $p$-paths is simply
$V^{p+1}$.  The following fact is useful.
\begin{lemma} \label{l:path-ind}
  Let $\eno\gamma n\in\Lambda_p$.  If in each of the paths there is an
  elementary $p$-path that is not in the other paths, then the
  $\gamma_i$'s are linearly independent.
\end{lemma}

\begin{definition}[Join of paths]
  The join of an elementary $p$-path $\gamma=(\eno[0]
  x p)$ and an elementary $q$-path $\sigma= (\eno[0]y q)$ is the
  elementary $(p+q+1)$-path $\gamma\sigma = (\eno[0]x p, \eno[0] y
  q)$.  For $\gamma=\sum_i a_i \gamma_i\in \Lambda_p$ and $\sigma =
  \sum_j b_j \sigma_j\in \Lambda_q$, where $\gamma_i$ and $\sigma_j$
  are elementary paths, their join is $\gamma \sigma = \sum_{ij}a_ib_j
  \gamma_i\sigma_j$.  On the other hand, if either
  $\sigma\in\Lambda_{-1}$ or $\gamma\in\Lambda_{-1}$, for example
  $\gamma = c\in\Reals$, then $\gamma\sigma = \sigma\gamma = c\sigma$,
  i.e., scalar multiplication by $c$.
\end{definition}
By the definition, the join is a tensor product.  Thus, the join can
be extended to multiple paths.  For example, if $\gamma$, $\sigma$,
and $\omega$ are paths, possibly in $\Lambda_{-1}$, then $\gamma\sigma
\omega = (\gamma\sigma)\omega = \gamma(\sigma\omega)$.

\begin{lemma}\label{l:join}
  Suppose $\eno\gamma n\in\Lambda_p$ are linearly independent, where
  $p\ge0$.  Let $\eno\sigma n\in\Lambda_q$, $q\ge-1$.  Then $\sum
  \gamma_i \sigma_i=0\Iff \seqop\sigma=n=0$.  Similarly,
  $\sum\sigma_i \gamma_i=0\Iff \seqop\sigma=n=0$.
\end{lemma}
\begin{proof}
  Since the $\gamma_i$'s are linearly independent, for each $1\le j\le
  n$, there is a linear function $f_j:\Lambda_p\to\Reals$, such that
  $f_j(\gamma_i)= \delta_{ij}$.  Let $g: \Lambda_q\to\Reals$ be an
  arbitrary linear function.  Then from the equivalence between tensor
  product and join, $(f_j\otimes g)(\sum_i\gamma_i\sigma_i) =
  \sum_i f_j(\gamma_i) g(\sigma_i)=g(\sigma_j)$.  If
  $\sum_i\gamma_i\sigma_i = 0$, then it follows that $g(\sigma_j)=0$.
  Since this holds for any linear function $g$, then $\sigma_j=0$.
  The other direction is clear.  The second part can be similarly
  proved.
\end{proof}

For $p\ge0$, the \emph{boundary\/} map $\partial_p: \Lambda_p\to
\Lambda_{p-1}$ is a linear map, such that for any elementary $p$-path
$\sigma=(\eno[0] x p)$ on $V$,
\[
  \partial_p\sigma =
  \begin{cases}
    1 & \text{if~} p=0\\
    \sum_{i=0}^p(-1)^ i (x_0,\ldots,\hat x_i , \ldots,x_p)
    &\text{if~} p>0,
  \end{cases}
\]
where $\hat x_i$ denotes deletion of $x_i$.  For $p\le-1$, by
$\Lambda_{p-1} = \{0\}$, define $\partial_p=0$.  From Lemma 2.6 of
\cite {grigoryan2012homologies}, it follows that for $\gamma\in
\Lambda_p$ and $\sigma\in\Lambda_q$, $p,q\ge-1$,
\begin{align} \label{e:bd-join}
  \partial_{p+q+1}(\gamma\sigma) =
  (\partial_p\gamma) \sigma + (-1)^{p+1}\gamma(\partial_{q}\sigma).
\end{align}
By $\partial_{p-1}\circ \partial_p =0$, $(\Lambda_p, \partial_p)_{p\in
  \Ints}$ is a chain complex \cite[Lemma 2.4]
{grigoryan2012homologies}.

The discussion so far only involves vertices.  We next take into
account edges.
\begin{definition}[Allowed paths on a digraph]
  Let $(\eno[0] x p)$ be an elementary $p$-path on $V$.  If either
  $p=0$ or $(x_i, x_{i+1})\in E$ for all
  $0\le i<p$, then the path is said to be \emph{allowed\/} on $G$.
  Let $\Cal A_p = \Cal A_p(G)$ be the vector space over $\Reals$
  generated by the set of allowed elementary $p$-paths on $G$.
  Elements in $\Cal A_p$ are called \emph{allowed $p$-paths\/}.
  Define $\Cal A_{-1}\equiv \Reals$ and $\Cal A_p\equiv\{0\}$ for
  $p<-1$.
\end{definition}

Path homologies of $G$ are defined on a chain complex of allowed
paths.  However, the natural choice $(\Cal A_p,
\partial_p)_{p\in\Ints}$ in general is not a chain complex
because the image of an allowed path under $\partial_p$ may not be
allowed.  This issue is resolved by the following.

\begin{definition}[Path homology group]
  The space of \emph{$\partial$-invariant $p$-paths\/} on $G$, denoted
  by $\Omega_p = \Omega_p (G)$, is the set of allowed $p$-paths
  $\gamma$ on $G$ with $\partial_p \gamma\in \Cal A_{p-1}(G)$.  From
  $\partial_p(\Omega_p) \subset\Omega_{p-1}$, we have a chain
  complex $\cdots \xrightarrow{\partial_3} \Omega_2
  \xrightarrow{\partial_2} \Omega_1 \xrightarrow{\partial_1} \Omega_0
  \xrightarrow{\partial_0} \Reals \xrightarrow{\partial_{-1}} 0$.  For
  each positive $p\in\Ints$, the $p$-dimensional (reduced)
  \emph{path homology group\/} of $G$ is
  \begin{align}\label{e:homology}
    H_p = H_p(G) =
    \frac{\ker(\partial_p|_{\Omega_p})}
    {\image(\partial_{p+1}|_{\Omega_{p+1}})}
    =\frac{\ker(\partial_p)\cap \Omega_p}
    {\partial_{p+1}(\Omega_{p+1})}
    =\frac{\ker(\partial_p)\cap \Cal A_p}
    {\partial_{p+1}(\Omega_{p+1})}.
  \end{align}
  Since $H_p$ is a vector space over $\Reals$, its dimension $\dim(H_p)$
  is called the $p$th (reduced) \emph{Betti number\/}, denoted
  $\beta_p(G)$ \citep {grigoryan2012homologies}.  Elements in
  $\ker(\partial_p)\cap \Cal A_p$ will be referred to as
  \emph{$p$-cycles\/},  and those in $\partial_{p+1}(\Omega_{p+1})$ as
  \emph{$p$-boundaries\/}.
\end{definition}

By this definition, a $p$-cycle is necessarily an allowed $p$-path.
Denote the maximum path length of $G$ by
\[
  \ell(G) = \max\{p\ge0: \Cal A_p\ne\{0\}\}.
\]
Finally, in analogy to allowed paths, for $p\ge0$, let $\Cal D_p =
\Cal D_p(G)$ be the vector space over $\Reals$ generated by the set of
elementary $p$-paths on $V$ that are not allowed on $G$.  Note that
$\Cal D_0=\{0\}$.  For $p<0$, define $\Cal D_p\equiv \{0\}$.  Then
$\Lambda_p = \Cal A_p\oplus \Cal D_p$.

\subsection{A general algorithm to compute path homology}
We recall the general algorithm of \cite {chowdhury2021path} that
computes the path homologies of a digraph.  Given two finite sets $S$
and $T$, denote $M \in \Reals^{S\times T}$ if $M$ is a real matrix
with rows indexed by $S$ and columns by $T$.  The null space of $M$ is
$\Cal N(M) = \{v\in \Reals^T: M v=0\}$ and its column space is $\Cal
C(M) =\{M v: v\in \Reals^T\}$.  By convention, if $S=\emptyset$ or
$T=\emptyset$, then $\Cal N(M)=\Reals^T$ and $\Cal C(M) = \{0\}$. 

Let $G=(V,E)$ be a digraph.  Since $H_p(G)=0$ for $p>\ell(G)$, we only
need to consider $0\le p\le \ell(G)$.  Recall the set of elementary
$p$-paths on $V$ is identical to $V^{p+1}$.  Denote by $A_p(G)$ the
set of allowed \emph{elementary\/} $p$-paths on $G$.  Note $A_0(G)=V$.
Let $N\Sp 0$ be the identity matrix in $\Reals^{V\times V}$ and
$B\Sp0\in \Reals^V$ the row vector of 1's.  Let $V = \{\eno{\omega\Sp
  0} {m_0}\}$.  Then, iteratively for $p=1,\ldots, \ell(G)$, do the
following steps.
\begin{enumerate}
\item Construct $A_p(G) = \{\eta x: \eta\in A_{p-1}(G), (u,x)\in E$
  with $u$ being the last vertex of $\eta\}$.
  
\item Compute matrix $D\Sp p\in \Reals^{V^p \times A_p(G)}$, such that
  for $\eta\in A_p(G)$, $\partial\eta = \sum_{\gamma \in V^p}\gamma
  D\Sp p_{\gamma\eta}$.
\item Construct a basis of $\Omega_p(G)$.  To do this, partition $D\Sp
  p$ as $\begin{pmatrix} D\Sp p_*\\ E\Sp p \end{pmatrix}$, such that 
  $D\Sp p_*$ consists of rows of $D\Sp p$ whose indices are in $V^p
  \setminus A_{p-1}(G)$, while $E\Sp p$ consists of those whose
  indices are in $A_{p-1}(G)$.  Let $N\Sp p$ be a matrix whose column
  vectors consist a basis of $\Cal N(D\Sp p_*)$.  Note that the rows 
  of $N\Sp p$ are indexed by $A_p(G)$.  Let $v_i = (v_{i,
    \eta})_{\eta\in A_p(G)}$, $i=1,\ldots, m_p$, be the column vectors 
  of $N\Sp p$.  Then $\omega\Sp p_i = \sum_{\eta\in A_p(G)} \eta
  v_{i,\eta}$, $1\le i\le m_p$, is a basis of $\Omega_p(G)$.
  
\item Generate an $m_{p-1}\times m_p$ matrix $B\Sp p = (B\Sp p_{ij})$,
  such that for each $1\le j\le m_p$, $\partial\omega\Sp p_j =
  \sum^{m_{p-1}}_{i=1} \omega\Sp{p-1}_i B\Sp p_{ij}$.  In fact, $B\Sp
  p$ can be any matrix satisfying $N\Sp{p-1} B\Sp p = E\Sp p N\Sp p$,
  which has at least one solution.
\end{enumerate}

Finally, for $p>\ell(G)$, define $B\Sp p=0$.  It can be shown that
$B\Sp p B\Sp{p+1}=0$ for $p\ge0$.  Then the homologies of $G$ of
dimension up to $\ell(G)$ can be calculated by $H_p(G)\cong \Cal 
N(B\Sp p)/\Cal C(B\Sp{p+1})$.

\section{Basic results} \label{s:basic}
Let $G=(V,E)$ be a digraph.  
\begin{definition}[Cross section and support of a path]
  \label{d:cs-support}
  Let $\alpha \in \Lambda_p(G)$, $p\ge0$.  Let $(x_{10}, \ldots,
  x_{1p})$, \ldots, $(x_{k0}, \ldots, x_{kp})$ be the elementary
  $p$-paths in $\alpha$.  For each $s=0,\ldots,p$, the cross section
  of $\alpha$ at depth $s$, denoted $[\alpha]_s$, is the set
  $\{x_{1s}, \ldots, x_{ks}\}$.  Define $[\alpha]_s=\emptyset$ if $s<0$
  or $s>p$.  The cross sections $[\alpha]_0$ and $[\alpha]_p$ are
  referred to as the top and the bottom of $\alpha$, denoted by
  $\Cil\alpha$ and $\Flr\alpha$, respectively.
  
  If in addition $\alpha\in\Cal A_p(G)$, then its support
  $\sppt(\alpha)$ is the subgraph of $G$ consisting of the vertices
  and edges in the elementary $p$-paths in $\alpha$, i.e.,
  $\sppt(\alpha) = (V_\alpha, E_\alpha)$ with $V_\alpha = \{x_{is}:
  1\le i\le k, 0\le s\le p\}$ and $E_\alpha= \{(x_{is}, x_{i, s+1}):
  1\le i\le k, 0\le s< p\}$. 
\end{definition}

By the definition, if $\alpha=0$, then $[\alpha]_s=\emptyset$ for all
$s$, and $\sppt(\alpha)=\emptyset$.  For $\alpha\in\Lambda_{-1}(G)
\equiv \Reals$, define $[\alpha]_s= \sppt(\alpha) = \emptyset$ for all
$s$.  The following simple fact is useful.
\begin{lemma} \label{l:cross-sec}
  Let $\eno\gamma n\in\Lambda_p(G)\setminus\{0\}$.  If for any two
  $i\ne j$, $[\gamma_i]_s\cap[\gamma_j]_s =\emptyset$ for at least one
  $0\le s\le p$, then all the $\gamma_i$'s are linearly independent.
\end{lemma}
\begin{proof}
  By the assumption, the sets of elementary paths of the $\gamma_i$'s
  are nonempty and disjoint from each other.  Then the proof follows
  from \Cref{l:path-ind}. 
\end{proof}

\begin{definition}[Predecessor and successor of a
  vertex]\label{d:pred-succ}
  For $x\in V$, denote 
  \[ 
    \pred(x)=\pred_G(x)=\{w \in V: (w,x) \in E\}, \quad
    \succ(x)=\succ_G(x)=\{y \in V: (x,y) \in E\}.
  \]
\end{definition}
It is not hard to see that for $\gamma\in \Lambda_p(G)$ and $\sigma\in
\Lambda_q(G)$, $\gamma\sigma\in \Cal A_{p+q+1}(G)\Iff$ the following
two conditions are satisfied: (1) $\gamma$ and $\sigma$ are allowed
paths, (2) $\Flr\gamma \subset\pred(x)$ for all $x\in\Cil\sigma$.
Note that (2) is equivalent to $\Cil\sigma\subset \succ(x)$ for all
$x\in\Flr\gamma$.  If $G$ is a DAG, then (2) holds if and only if
$\Flr\gamma$, $\Cil\sigma$, and the edges in between consist a fully
connected stratified digraph of depth 1, commonly known as a bipartite
digraph, with $\Flr\gamma$ being the $0$th layer, and $\Cil\sigma$ the
$1$st layer.

Our algorithm relies on certain decomposition of cycles, which in turn
is based on certain decomposition of allowed paths.  Recall that a
cycle must be an allowed path.  The following result is an extension
of Proposition 4 of \cite{chowdhury2019path}.
\begin{proposition}[Decomposition of allowed paths and
  cycles]\label{p:decomp}
  Let $\alpha\in\Cal A_p(G)\setminus\{0\}$.  Then for each $s=0,
  \ldots, p$, $\alpha$ can be written as
  \begin{align} \label{e:decomp}
    \alpha=\sum_{x\in [\alpha]_s} \sum^{n_x}_{i=1}
    \gamma_{xi} x \sigma_{xi},
  \end{align}
  where for each $x\in[\alpha]_s$, $n_x\ge1$, $\gamma_{x1}, \ldots,
  \gamma_{xn_x}$ are linearly independent allowed $(s-1)$-paths with
  $\Flr{\gamma_{xi}} \subset\pred(x)$, and $\sigma_{x1}, \ldots,
  \sigma_{xn_x}$ are linearly independent allowed $(p-s-1)$-paths with 
  $\Cil{\sigma_{xi}}\subset\succ(x)$.

  Furthermore, if $\alpha$ is a $p$-cycle and $[\alpha]_{s-1}\cap
  [\alpha]_s = [\alpha]_s \cap [\alpha]_{s+1} = \emptyset$, then in
  the decomposition \eqref{e:decomp}, $\gamma_{xi}$ and $\sigma_{xi}$
  are cycles satisfying $\sum_{x\in [\alpha]_s} \sum^{n_x}_{i=1}
  \gamma_{xi} \sigma_{xi}=0$.
\end{proposition}

\begin{remark}
  The condition $[\alpha]_{s\pm1} \cap [\alpha]_s =\emptyset$, which
  is used the decomposition of a $p$-cycle, does not hold for 
  $p$-paths in general.  For example, if
  $(x_0,x_1,x_2,x_3)$ is an allowed elementary 4-path, then $\alpha =
  (x_0,x_1,x_2) + (x_1,x_2,x_3)$ is an allowed 3-path with
  $[\alpha]_1\cap [\alpha]_2 = \{x_1\}$.  On the other hand, if $G$ is
  a stratified digraph, and if $\Flr\alpha$ is in a single layer
  $K_i$, then $[\alpha]_s \subset K_{i+s-p}$ for each $0\le s\le p$,
  so the cross sections are disjoint.
\end{remark}

\begin{proof}
  If $p=0$, then $\alpha = \sum_{x\in [\alpha]_0} c_x x$ with $c_x
  \ne0$ being real numbers.  Meanwhile, $s$ can only be 0 in \eqref
  {e:decomp}.  Letting $n_x=1$, $\gamma_{1x}=c_x$, and $\sigma_{1x}=1$
  then proves the result.  Let $p\ge1$ from now on.  By grouping the
  elementary $p$-paths in $\alpha$ according to their $s$th vertices,
  $\alpha$ can be uniquely written as $\sum_{x\in [\alpha]_s}
  \alpha_x$, where $\alpha_x$ is an allowed $p$-path with
  $[\alpha_x]_s = \{x\}$.  Each $\alpha_x$ can be written as $\sum_i
  \omega_{xi} x  \eta_{xi}$, where each $\omega_{xi} x \eta_{xi}$ is
  an allowed $p$-path.  Among all such decompositions of $\alpha_x$,
  there is one with the smallest number of terms in the sum.  Let this
  decomposition of $\alpha_x$ be $\sum^{n_x}_{i=1} \gamma_{xi} x
  \sigma_{xi}$.  Since $\alpha_x\ne0$, $n_x\ge1$.  Assume that one of
  the $\gamma_{xi}$'s is a linear combination of the others, say,
  $\gamma_{x n_x} = \sum^{n_x-1}_{i=1} c_i \gamma_{xi}$.  Then
  \[
    \alpha_x = \sum^{n_x-1}_{i=1} \gamma_{xi}x\sigma_{xi} +
    \Grp{\sum^{n_x-1}_{i=1} c_i \gamma_{xi}} x\sigma_{xn_i}
    = \sum^{n_x-1}_{i=1} \gamma_{xi}x \tilde \sigma_{xi}, 
  \]
  where $\tilde\sigma_{xi} = \sigma_{xi} + c_i \sigma_{xn_i}$.  Since
  each $\gamma_{xi} x \sigma_{xi}$ is an allowed $p$-path, from the
  comment after \Cref{d:pred-succ},  $\Cil{\sigma_{xi}}\subset
  \succ(x)$.  As a result, $\Cil{\tilde\sigma_{xi}}\subset\succ(x)$
  and so $\gamma_{xi} x \tilde \sigma_{xi}$ is an allowed $p$-path.
  Then $\alpha_x$ has a decomposition with fewer than $n_x$ terms,
  which is a contradiction.  Thus the $\gamma_{xi}$'s are linearly
  independent.  Likewise, the $\sigma_{xi}$'s are linearly
  independent.

  Now suppose $\alpha$ is a $p$-cycle and $[\alpha]_{s\pm 1} \cap
  [\alpha]_s = \emptyset$.  From \eqref{e:bd-join} and
  \eqref{e:decomp},
  \[
    \partial_p \alpha=
    \underbrace{
      \sum_{x\in [\alpha]_s} \sum^{n_x}_{i=1}
      [\partial_{s-1} (\gamma_{xi} x)] \sigma_{xi}
    }_{\omega_1} +\underbrace{
      (-1)^{s+1}
      \sum_{x\in [\alpha]_s} \sum^{n_x}_{i=1}
      \gamma_{xi}x(\partial_{p-s-1}\sigma_{xi})
    }_{\omega_2}.
  \]
  By $\partial_p\alpha=0$, $\omega_1 + \omega_2=0$.  If $s=p$, it is
  clear that $\partial_{p-s-1}\sigma_{xi}=0$, giving $\omega_2=0$ and
  so $\omega_1=0$.  Let $s\le p-1$.  It is seen that $[\omega_1]_s
  \subset\bigcup_{x,i}\Cil{\sigma_{xi}} \subset [\alpha]_{s+1}$,
  while $[\omega_2]_s\subset [\alpha]_s$.  Since $[\alpha]_s
  \cap[\alpha]_{s+1}=\emptyset$, then $[\omega_1]_s \cap
  [\omega_2]_s=\emptyset$.  Since both $\omega_1$ and $\omega_2$ are 
  $(p-1)$-paths and $\omega_1+\omega_2=0$, from \Cref{l:cross-sec},
  both $\omega_1$ and $\omega_2$ are 0.  With similar argument, all
  the nonzero elements in the set $\{\omega_{2x} := \sum^{n_x}_{i=1}
  \gamma_{xi} x(\partial_{p-s-1}\sigma_{xi})$, $x\in[\alpha]_s\}$ are
  linearly independent, so by $\omega_2 = \sum_{x\in[\alpha]_s}
  \omega_{2x}$, each $\omega_{2x}=0$.  Given $x\in[\alpha]_s$, since
  $\gamma_{xi}$, $i=1,\ldots, n_x$ are linearly independent, then by
  applying \Cref{l:join} to $\omega_{2x}$, $x(\partial_{p-s-1}
  \sigma_{xi})=0$, giving $\partial_{p-s-1}\sigma_{xi}=0$.  Therefore,
  all $\sigma_{xi}$ are $(p-s-1)$-cycles.  Likewise, all $\gamma_{xi}$
  are $(s-1)$-cycles.  Then $\omega_1 = (-1)^s \sum_{x\in[\alpha]_s}
  \sum^{n_x}_{i=1} \gamma_{xi} \sigma_{xi}$.  By $\omega_1=0$, the
  proof is complete.
\end{proof}

In \Cref{p:decomp}, if $s=0$, then \eqref{e:decomp} becomes $\alpha =
\sum_{x\in\Cil\alpha} x \sigma_x$, where $\sigma_x\in \Cal A_{p-1}(G)
\setminus \{0\}$ with $\Cil{\sigma_x}\subset \succ(x)$, and if
$\alpha$ is a cycle with $\Cil\alpha\cap [\alpha]_1=\emptyset$, then
$\sigma_x$ are cycles and $\sum_{x\in\Cil\alpha}\sigma_x=0$.
Likewise, if $s=p$, \eqref {e:decomp} becomes $\alpha =\sum_{x\in
  \Flr\alpha} \gamma_x x$, where $\gamma_x\in \Cal A_{p-1}(G)
\setminus\{0\}$ with $\Flr{\gamma_x}\subset \pred(x)$, and if $\alpha$
is a cycle with $\Flr\alpha\cap [\alpha]_{p-1} = \emptyset$, then
$\gamma_x$ are cycles and $\sum_{x\in \Flr\alpha} \gamma_x = 0$.  The
following is a schematic of a cycle with decomposition
$\sum^4_{i=1}\gamma_i x_i$,
\[
  \begin{tikzpicture}
    \draw (0,1) rectangle (2,2);
    \draw (0,1) -- (1,.5) -- (2,1);
    \draw (1,.2) node {\scriptsize$x_1$};
    \draw (1,1.5) node {\scriptsize$\gamma_1$};
    
    \draw (1.5,1) rectangle (3.7,2);
    \draw (1.5,1) -- (2.6,.5) -- (3.7,1);
    \draw (2.6,.2) node {\scriptsize$x_2$};
    \draw (2.6,1.5) node {\scriptsize$\gamma_2$};
    
    \draw (3.4,1) rectangle (4.6,2);
    \draw (3.4,1) -- (4,.5) -- (4.6,1);
    \draw (4, .2) node {\scriptsize$x_3$};
    \draw (4,1.5) node {\scriptsize$\gamma_3$};
    \draw (4.2,1) rectangle (6,2);
    \draw (4.2,1) -- (5.1,.5) -- (6,1);
    \draw (5, .2) node {\scriptsize$x_4$};
    \draw (5,1.5) node {\scriptsize$\gamma_4$};
  \end{tikzpicture}
\]
where $x_i$ are different vertices in the same layer.  To express the
condition $\Flr{\gamma_i} \subset \pred(x_i)$, $\gamma_i$ is drawn as
a rectangle and $x_i$ as the tip of a triangle flipped upside down,
such that the rectangle and the triangle share a common bottom.  Note
that the $\gamma_i$'s share allowed elementary paths as their sum is
0.  In the schematic, this is shown as overlapping rectangles.  As a
concrete example, suppose $\gamma_1 = a_2 - a_1$, $\gamma_2 = 2 a_3 -
a_1 - a_2$, $\gamma_3 = a_4 - a_3$, but $\gamma_4$ is not given, where
$a_i$ are vertices in the same layer.  It is easy to check
$\partial\gamma_i = 0$ for $i=1,2,3$.  Since $\sum^4_{i=1} \gamma_i
x_i$ is a cycle, then $\gamma_4 = -(\gamma_1 + \gamma_2 + \gamma_3) =
2 a_2 - a_3 - a_4$.

\begin{corollary}\label{c:decomp}
  Let $x\in V$.
  \begin{enumerate}
  \item\label{c:decomp-pred} 
    If $|\pred(x)|\le1$, then for any cycle $c$ on $G$ with disjoint
    cross sections, either $x\not\in\sppt(c)$ or $x\in\Cil c$.
  \item \label{c:decomp-succ}
    If $|\succ(x)|\le1$, then for any cycle $c$ on $G$ with disjoint
    cross sections, either $x\not\in\sppt(c)$ or $x\in \Flr c$.
  \end{enumerate}
\end{corollary}
\begin{proof}
  We only show \ref{c:decomp-pred}.  The proof for \ref{c:decomp-succ} is
  similar.  Let $c$ be a $p$-cycle.  For $p=0$, by $\sppt(c)=\Cil c$,
  the result is trivial.  Let $p\ge1$ and $x\in \sppt(c)$.  Then
  $c\ne0$.  We need to show $x\in\Cil c$.  If $|\pred(x)|=0$, by
  \Cref{d:cs-support,d:pred-succ}, it is clear that $x\in\Cil c$.
  Let $|\pred(x)|=1$.  Then there is a unique $u\in V$ such that $(u,
  x)\in  E$.   Assume that $x\in [c]_s$ for some $0<s\le p$.  Since by
  assumption all the cross sections of $c$ are disjoint, by
  \Cref{p:decomp}, we can write $c=\sum_{i=1}^{n_x} \gamma_{x i} x
  \sigma_{x i}+\sum_{y\in [c]_s \backslash\{x\}} \sum_{i=1}^{n_y}
  \gamma_{y i} y \sigma_{y i}$, such that $n_x>0$ and each $\gamma_{x
    i}$ is an $(s-1)$-cycle with $\Flr{\gamma_{xi}} \subset \pred(x) =
  \{u\}$.  As a result, $\gamma_{xi} = \tilde\gamma_{xi} u$ for some
  $(s-2)$-path $\tilde\gamma_{xi}$.  However, applying \Cref{p:decomp}
  to $\gamma_{xi}$, it is seen that $\tilde\gamma_{xi}=0$, giving
  $\gamma_{xi}=0$.  The contradiction shows that $s$ has to be 0,
  i.e., $x\in \Cil c$.
\end{proof}

\Cref{c:decomp} leads to a pruning procedure that helps reduce
computation of the full-depth path homology.  Let $G$ be a stratified
digraph with layers $\eno[0] K L$.  If $x$ is a vertex of $G$ such
that $x\not\in K_0$ and $|\pred(x)|\le1$ or $x\not\in K_L$ and
$|\succ(x)|\le1$, then by \Cref{c:decomp}, $x$ cannot be in any
full-depth cycle of $G$, so it can be deleted without changing the
space of full-depth cycles.  For this reason, $x$ is said to be
removable.  After $x$ and the edges to or from it are deleted from
$G$, the resulting digraph $G_1$ is still stratified and has the same
space of full-depth cycles as $G$.  Note that due to the deletion of
edges, some vertices that are not removable from $G$ may become
removable from $G_1$.  We can then delete any such vertex from $G_1$
and iterate until no vertex is removable.  See \Cref{s:pruning} for
more detail.

As the next result suggests, further pruning can be done, possibly at
a higher computational cost.  Its proof follows the one for \Cref
{c:decomp} and so is omitted for brevity.  Two vertices are said to be
\emph{connected\/} if there is an allowed elementary path that
contains both.
\begin{corollary} \label{c:decomp2}
  Let $G$ be a stratified digraph with layers $\eno[0] K L$ and $x\in
  K_i$.  If there is $j\ne i$, such that $|\{u\in K_j: u$ and $x$ are
  connected$\}|\le1$, then $x$ is not in any full-depth cycle of $G$.
\end{corollary}

Besides pruning, another way that helps reduce computation is to
partition $G$ into \emph{weakly connected components\/}, which
are non-empty subgraphs components $\eno G m$ whose vertex sets and
edge sets partition the vertex set and edge set of $G$, respectively,
with each $G_i$ being connected when treated as an undirected graph.
The weakly connected components can be found in time $O(|V|+|E|)$
\citep {sedgewick2011algorithms, pacault1974computing}.  Since $H_l(G)
= \oplus^m_{i=0} H_l(G_i)$ holds for any $l$, the computation of the
$l$-dimensional homology of $G$ can be reduced to that for the
individual smaller components.

The decomposition of cycles in \Cref{p:decomp} has a consequence for topological invariance of the full-depth path homology of a stratified digraph.  By standard argument, the homology is invariant under graph isomorphisms.  However, it is easy to see that it is not invariant under operations such as edge or vertex contractions.  As an extreme example, if all the vertices in a layer are merged into a single vertex, then from \Cref{c:decomp}, the full-depth path homology degenerates into 0.  A question is, under what graph operations is full-depth path homology invariant?  The next result implies that, if a stratified digraph is made one layer deeper with the new bottom layer consisting of two vertices that are connected to every vertex in its original bottom layer, then its full-depth path homology is invariant in the sense that it is isomorphic that of the original graph, although its dimension is higher by 1.
\begin{corollary} \label{c:invariance}
  Let $G=(V,E)$ and $G'=(V', E')$ be two stratified digraphs, such that $G$ has layers $\eno[0] KL$, $G'$ has layers $\eno[0] KL, K_{L+1}$, and $E' = E\cup \{(u,v): u\in K_L, v\in K_{L+1}\}$.  Then $H_{L+1}(G')\cong [H_L(G)]^{|K_{L+1}|-1}$, i.e., the $(|K_{L+1}|-1)$-fold direct sum of $H_L(G)$.  
\end{corollary}
\begin{proof}
  Let $K_{L+1}=\{\eno[0] x n\}$.  By $\Cil{x_i} = K_L$, for each $L$-cycle $\gamma$ in $G$, $\gamma x_i$ is an allowed path in $G'$.  Then, for any $L$-cycles  $\eno\gamma n$ in $G$, $f(\eno x n):=-(\cum \gamma n) x_0 + \sum^n_{i=1} \gamma_i x_i$ is an $(L+1)$-cycle in $G'$.  Conversely,
  from the above discussion, any $(L+1)$-cycle in $G'$ can be uniquely written as $\sum^n_{i=0} \gamma_i x_i$, where $\gamma_i$ are $L$-cycles in $G'$ and $\gamma_0 = -(\cum \gamma n)$.  Note that $\gamma_i$ are also $L$-cycles in $G$.  It follows that $f(\eno \gamma n)$ is a linear bijection from $[H_L(G)]^n$ to $H_{L+1}(G')$.
\end{proof}

Now let $G$ be a DAG.  Consider the maximal path homology of $G$.
When $\ell(G)=0$, $G$ only has isolated vertices, so $\beta_0(G) =
|V|-1$ \cite[cf.][]{chowdhury2019path}.  For $\ell(G)\ge1$, the next
result shows that the calculation of the maximal path homology of $G$
can be reduced to that for full-depth path homologies of stratified
subgraphs of $G$.
\begin{proposition} \label{p:subgraph}
  Let $G$ be a DAG with $\ell(G)>0$.  Let $G_*=\cup_{\gamma \in \Cal
    A_{\ell(G)}} \sppt(\gamma)$ and $K_i = \cup_{\gamma \in \Cal
    A_{\ell(G)}} [\gamma]_i$, $i=0,\ldots,\ell(G)$.  Then
  \begin{enumerate}
  \item \label{p:subgraph1}
    $G_*$ is a stratified digraph with layers $\eno[0] K {\ell(G)}$,
  \item \label{p:subgraph2}
    every allowed elementary path in $G_*$ can be extended into an
    allowed elementary $\ell(G)$-path in $G_*$, and
  \item \label{p:subgraph3}
    $H_{\ell(G)}(G) = H_{\ell(G)}(G_*)$.
  \end{enumerate}
\end{proposition}
\begin{remark}
  The longest path length of a DAG and its weakly connected components
  can be computed efficiently using available software.  The relevant
  information is given in \Cref{ss:auxiliary}.  Following the method
  to compute $l(G)$, $G_*$ can also be computed efficiently.  Since we
  have not been able to find an existing algorithm to compute $G_*$,
  for completeness, an algorithm is provided in \Cref{ss:auxiliary}.
\end{remark}
\begin{proof}
  By construction, all the allowed $\ell(G)$-path in $G$ are in
  $G_*$, and every vertex or edge must be in an  allowed elementary
  $\ell(G)$-path in $G_*$.  We claim that every vertex $x$ in $G_*$
  is at the same position in every allowed elementary $\ell(G)$-path
  that it is in.  Indeed, any two such paths can be written
  as $\gamma_1  x\sigma_1$ and $\gamma_2 x\sigma_2$, where $\gamma_k$
  is an allowed elementary $i_k$-path and $\sigma_k$ an allowed
  elementary $[\ell(G)-2 - i_k]$-path with $-1\le i_k<\ell(G)$.  Since
  $\gamma_1 x\sigma_2$ is 
  an allowed $[\ell(G)+i_1 -i_2]$-path and $\gamma_2 x\sigma_1$ an
  allowed $[\ell(G)+i_2-i_1]$-path, and each has length no greater
  than $\ell(G)$, then $i_1=i_2$, hence the claim.

  \ref{p:subgraph1}\
  From the above argument, every vertex in $G_*$ is in a unique
  $K_i$.  Thus the $K_i$'s are disjoint.  If $(x,y)$ is an edge in
  $G_*$, then it is in an allowed elementary $\ell(G)$-path.  If $x$
  is at the $i$th position in the path, then $y$ is at the $(i+1)$st
  position.  In other words, if $(x,y)$ is an edge in $G_*$, then
  there is $i$ such that $x\in K_i$ and $y\in K_{i+1}$.  This shows
  that $G_*$ is a stratified digraph and the $K_i$'s are its layers.
  
  \ref{p:subgraph2}\  Let $\eta$ be an allowed elementary $p$-path in
  $G_*$.  Let $p<\ell(G)$, for otherwise there is nothing to show.  If
  $p\le1$, then $\eta$ is a vertex or edge, a case already covered in
  the starting argument.  Let $p\ge2$ and suppose every allowed
  elementary $(p-1)$-path in $G_*$ can be extended into an allowed
  elementary $\ell(G)$-path in $G_*$.  Let $\eta=(\eno[0] u p)$.  Then
  $(u_0, u_1)$ is in an allowed elementary $\ell(G)$-path $\eta_1=
  \gamma_1(u_0,u_1)\sigma_1$.  On the other hand, by the hypothesis,
  $(\eno u p)$ is in an allowed elementary $\ell(G)$-path $\eta_2=
  \gamma_2(\eno u p)\sigma_2$.  From the starting argument, $u_1$ is
  at the same position in $\eta_1$ and $\eta_2$.  Then it is easy to
  see that $\gamma_1\eta\sigma_2$ is an allowed elementary
  $\ell(G)$-path in $G_*$.
  
  \ref{p:subgraph3}
  Since the space of $\ell(G)$-boundaries is trivial for $G$ and
  $G_*$, and $\Cal A_{\ell(G)}(G) = \Cal A_{\ell(G)}(G_*)$, then
  $H_{\ell(G)}(G) = \ker(\partial_{\ell(G)})\cap\Cal A_{\ell(G)}(G)
  =\ker(\partial_{\ell(G)})\cap \Cal A_{\ell(G)}(G_*)
  =H_{\ell(G)}(G_*)$.
\end{proof}

The last result of this section pertains to $\partial$-invariant
paths.  For $p\le1$, it is plain that $\Omega_p = \Cal A_p$, while
generally speaking, for $p\ge2$, $\Omega_p$ is hard to
characterize.  However, for a stratified digraph, the following result
holds.  Although the result is not used in the rest part of the paper,
it may be of interest in its own right.

\begin{proposition}[Decomposition of $\partial$-invariant paths in a
  stratified digraph]
  \label{p:decomp3}
  Suppose $G=(V,E)$ is a stratified digraph with layers $\eno[0] K L$.
  Let $2\le p\le L$ and $\sigma\in\Cal A_p(G)$.  Then $\sigma\in
  \Omega_p(G)$ if and only if $\sigma$ can be written as
  $\sum x\gamma_{x,y} y$, where the sum is over all $x\in K_i$, $y\in
  K_{i+p}$, and $0\le i\le L-p$, and for each pair $(x,y)$ in the
  sum, $\gamma_{x,y}$ is a $(p-2)$-cycle with $\Cil{\gamma_{x,y}}
  \subset \succ(x)$ and $\Flr{\gamma_{x,y}}\subset \pred(y)$.   As
  a result, $\Omega_p(G) \cong \bigoplus E_{x,y}$, where $E_{x,y} =
  \{\gamma\in \ker(\partial) \cap \Cal A_{p-2}(G): x\gamma y\in\Cal
  A_p(G)\}$.
\end{proposition}
\begin{proof}
  Let $p\ge2$ and $\sigma \in\Cal A_p(G)$.  By grouping the 
  elementary paths in $\sigma$ according to their initial and last
  vertices, $\sigma = \sum x \gamma_{x,y} y$, where the sum is over
  all $x\in K_i$, $y\in K_{i+p}$, and $0\le i\le L-p$, for each pair
  $(x,y)$ in the sum, $\gamma_{x,y} \in \Cal A_{p-2}(G)$ such that
  $x\gamma_{x,y}y\in \Cal A_p(G)$.  Then $\partial\sigma = \sum
  [\gamma_{x,y} y - x(\partial \gamma_{x,y}) y + (-1)^p
  x\gamma_{x,y}]$.

  Suppose $\sigma\in\Omega_p(G)$.  Then by definition, $\partial\sigma
  \in\Cal A_{p-1}(G)$.  Since $\gamma_{x,y} y$ and $x\gamma_{x,y}\in
  \Cal A_{p-1}(G)$, then $\sum x(\partial\gamma_{x,y}) y\in \Cal
  A_{p-1}(G)$.  However, for any allowed elementary $p$-path $x\omega
  y\in \Cal A_p(G)\setminus\{0\}$ with $p\ge2$, it is easy to see
  that all the elementary $(p-1)$-paths in $x(\partial\omega) y$ are
  not allowed.
  It follows that $\sum x(\partial\gamma_{x,y}) y \in \Cal
  D_{p-1}(G)$.  By $\Lambda_p(G)=\Cal A_p(G) \oplus \Cal D_p(G)$, $\sum
  x(\partial\gamma_{x,y}) y=0$.  From \Cref{l:cross-sec}, all the nonzero
  $(p-1)$-paths $x(\partial\gamma_{x,y})y$ with different $(x,y)$ are
  linearly independent.  As a result, for every $(x,y)$,
  $x(\partial\gamma_{x,y})y=0$, and hence $\partial\gamma_{x,y}=0$.
  Then $\gamma_{x,y}$ is a $(p-2)$-cycle.
 
  On the other hand, suppose for every pair $(x,y)$ in the sum,
  $\gamma_{x,y}$ is a $(p-2)$-cycle and $x\gamma_{x,y}y\in \Cal
  A_p(G)$.  Then $\partial(x\gamma_{x,y} y) = (-1)^p x\gamma_{x,y} +
  \gamma_{x,y} y\in \Cal A_{p-1}(G)$.  Then by definition,
  $x\gamma_{x,y} y\in \Omega_p(G)$.  As a result, $\sigma \in
  \Omega_p(G)$.
\end{proof}

\section{Full-depth homology of stratified digraph} \label{s:recursive}
In this section, let $G=(V,E)$ be a stratified digraph with layers
$\eno[0] K L$.  Since the space of $L$-boundaries of $G$ is trivial,
$H_L(G) = \ker(\partial_L)$.  Fix $0\le p\le L$.

\begin{definition} \label{d:G_S}
  For $S\subset K_p$, denote by $G_S$ the union of
  $\sppt(\gamma)$ over all $\gamma \in \Cal A_p(G)$ with
  $\Flr\gamma\subset S$.  For brevity, if $S=K_p$, then write $G_p 
  = G_{K_p}$.
\end{definition}

\begin{proposition}[Decomposition of a cycle in a stratified
  digraph]\label{p:cycle-sd}
  For $x\in K_p$, denote
  \[
    \Cal C^x_{p-1}
    =\ker(\partial_{p-1})\cap\Cal A_{p-1}(G_{\pred(x)}),
  \]
  and for  $S\subseteq K_p$, let
  \begin{align} \label{e:S+}
    S^+ = \{x\in S: \Cal C^x_{p-1}\ne\{0\}\}.
  \end{align}
  Every $\alpha\in\ker(\partial_p)\cap\Cal A_p(G_S)$ has a unique
  decomposition $\sum_{x\in S^+} \gamma_x x$ such that $\gamma_x\in
  \Cal C^x_{p-1}$ and $\sum_{x \in S^+} \gamma_x=0$.  The map $\pi_S:
  \alpha \mapsto(\gamma_x)_{x\in S^+}$
  is a linear bijection from $\ker(\partial_p) \cap \Cal A_p(G_S)$ to
  \[
    \Cal V^*_S = \Cbr{(\gamma_x)_{x \in S^+} \in \bigoplus_{x \in
        S^+}\Cal C^x_{p-1}: \sum_{x \in S^+} \gamma_x=0},
  \]
  where $(\gamma_x)_{x\in S^+}:=0$ and $\Cal V^*_S:=\{0\}$ if
  $S^+=\emptyset$.  Furthermore, for any $S\subseteq T\subseteq K_p$,
  the following commutative diagram holds
  \[ 
    \begin{tikzcd}
      \ker(\partial_p)\cap \Cal A_p(G_S)\arrow{r}{\pi_S} 
      \arrow[hook,swap]{d}{\imath} & \Cal V^*_S \arrow[hook]{d}\\      
      \ker(\partial_p)\cap \Cal A_p(G_T)\arrow{r}{\pi_T} & \Cal
      V^*_T
    \end{tikzcd}
  \]
  where $\imath: ker(\partial_p)\cap \Cal A_p(G_S)
  \hookrightarrow\ker(\partial_p)\cap \Cal A_p(G_T)$ is the inclusion
  map, and $\Cal V^*_S\hookrightarrow \Cal V^*_T$ is the natural
  embedding $(\gamma_x)_{x \in S^+}\mapsto (\tilde\gamma_x)_{x\in
    T^+}$ such that $\tilde\gamma_x = \gamma_x$ for $x\in S^+$ and
  $\tilde\gamma_x=0$ for $x\in T^+\setminus S^+$.
\end{proposition}
\begin{remark}
  When $p=0$, by $\ker(\partial_{-1})=\Cal A_{-1}(G)=\Reals$ for any
  digraph $G$, including the empty one, $\Cal C^x_{-1}=\Reals$ for
  $x\in K_0$, and so $S^+=S$ for any $S\subseteq K_0$.
\end{remark}
\begin{proof}
  Since $G_S$ is a stratified digraph and $\alpha\in\ker(\partial_p)
  \cap\Cal A_p(G_S)$, the cross sections of $\alpha$ are disjoint.
  From \Cref{p:decomp} and the remark that follows, $\alpha$ can be
  written as $\sum_{x\in\Flr\alpha} \gamma_x x$, such that
  $\sum_{x\in \Flr\alpha} \gamma_x=0$ and for each $x\in\Flr\alpha$,
  $\gamma_x\in \ker(\partial_{p-1})\setminus\{0\}$ with $\Flr{\gamma_x}
  \subset \pred(x)$, i.e., $\gamma_x\in \Cal C^x_{p-1}\setminus\{0\}$.
  Then $\Flr\alpha\subset S^+$.  Let $\gamma_x=0$ for $x\in
  S^+\setminus\Flr\alpha$.  Then $\alpha= \sum_{x\in S^+} \gamma_x x$
  with $\gamma_x\in \Cal C^x_{p-1}$ and $\sum_{x\in S^+} \gamma_x =0$.
  From \Cref{l:join}, the decomposition is unique, which yields that
  $\pi_S$ is a linear map from $\ker(\partial_p)\cap \Cal A_p(G_S)$ to
  $\Cal V^*_S$.  It is clear that $\pi_S$ is 1-to-1.  On the other
  hand, for any $(\gamma_x)_{x\in S^+}\in \Cal V^*_S$, it is easy to
  check that $\alpha:=\sum_{x\in S^+}\gamma_x x\in \Cal A_p(G_S)$ and
  $\partial_p\alpha=0$.  Thus $\pi_S$ is bijective.  From the
  uniqueness of the decomposition, the commutative diagram easily
  follows.
\end{proof}

\begin{theorem}[Cycle space recursion]\label{t:cycle}
  Let $\{\eno\gamma N\}$ be a basis of $\ker(\partial_{p-1}) \cap \Cal
  A_{p-1}(G_{p-1})$.  For $x\in K^+_p$, where $K^+_p$ is as in
  \eqref{e:S+}, let $\eno{\gamma^x}{N_x}$ form a basis of $\Cal 
  C^x_{p-1}$.  Let $A_x$ be the matrix such that
  $(\eno{\gamma^x}{N_x}) = (\eno\gamma N) A_x$ and $h_x: \Cal
  C^x_{p-1}\to \Reals^{N_x}$ the linear map such that for $\gamma\in
  \Cal C^x_{p-1}$, $\gamma = (\eno{\gamma^x}{N_x}) h_x(\gamma)$.  For
  $S\subseteq K_p$, let $h_S = \oplus_{x\in S^+} h_x$ and $\pi_S$ be
  as in \Cref{p:cycle-sd}, where $h_S:=0$ and $\pi_S:=0$ if
  $S^+=\emptyset$.   Then $\phi_S = h_S\circ\pi_S$ is a linear
  bijection from $\ker(\partial_p) \cap \Cal A_p(G_S)$ to
  \[
    \Cal V_S
    :=\Cbr{(v_x)_{x \in S^+} \in \bigoplus_{x \in S^+} \Reals^{N_x}:
      \sum_{x \in S^+} A_x v_x=0}
  \]
  with $\Cal V_S:=\{0\}$ if $S^+ = \emptyset$, and for any $S\subseteq
  T\subseteq  K_p$, the following commutative diagram holds
  \[
    \begin{tikzcd}
      \ker(\partial_p)\cap \Cal A_p(G_S)\arrow{r}{\phi_S}
      \arrow[hook,swap]{d}{\imath} & \Cal V_S \arrow[hook]{d}\\
      \ker(\partial_p)\cap \Cal A_p(G_T)\arrow{r}{\phi_T} & \Cal
      V_T,
    \end{tikzcd}
  \]
  where $\Cal V_S\hookrightarrow \Cal V_T$ is the natural embedding 
  $(v_x)_{x \in S^+}\mapsto (\tilde v_x)_{x\in T^+}$ such that
  $\tilde v_x = v_x$ for $x\in S^+$ and $\tilde v_x=0$ for $x\in
  T^+\setminus S^+$.
\end{theorem}
\begin{proof}
  From \Cref{p:cycle-sd}, it suffices to show that $h_S$ is a linear
  bijection from $\Cal V^*_S$ to $\Cal V_S$ such that for $S\subset
  T\subset K_p$, the following commutative diagram holds
  \[
    \begin{tikzcd}
      \Cal V^*_S \arrow[r, "h_S"] 
      \arrow[d, hook] & \Cal V_S \arrow[d, hook]\\
      \Cal V^*_T\arrow[r, "h_T"] & \Cal V_T
    \end{tikzcd}
  \]
  Let $(\gamma_x)_{x\in S^+}\in \Cal V^*_S$.  Then for each $x\in
  S^+$, by $\gamma_x\in \Cal C^x_{p-1}$, $\gamma_x = (\eno{\gamma^x}
  {N_x}) v_x$ with $v_x =  h_x(\gamma_x)$.  Next,
  $\sum_{x\in S^+}\gamma_x= \sum_{x\in S^+} (\eno{\gamma^x}{N_x}) v_x
  =(\eno\gamma N) \sum_{x\in S^+} A_x v_x$.  Because $\sum_{x\in S^+}
  \gamma_x=0$,  $\sum_{x\in S^+} A_x v_x = 0$.  Then
  $h_S((\gamma_x)_{x\in S^+}) = (h_x(\gamma_x))_{x\in S^+} =
  (v_x)_{x\in S^+}\in \Cal V_S$.  On the other hand, for $(v_x)_{x\in
    S^+}\in \Cal V_S$, let $\gamma_x = (\eno{\gamma^x} {N_x}) 
  v_x$ for each $x\in S^+$.  Then $\gamma_x \in\Cal C^x_{p-1}$ and
  $\sum_{x\in S^+} \gamma_x = (\eno\gamma N)\sum_{x\in S^+} A_x
  v_x=0$.  Therefore, $h_S$ is a bijection from $\Cal V^*_S$ to $\Cal
  V_S$.  Finally, it is plain that the above commutative diagram
  holds.
\end{proof}

\section{Algorithms} \label{s:algorithms}
\subsection{Recursive calculation of full-depth homology}
Let $G$ be a stratified digraph with layers $\eno[0] K L$.  Recall that for $0\le p\le L$, $G_p$ is the subgraph of $G$ consisting of layers $\eno[0] K p$ and all the edges of $G$ between those layers; see \Cref{d:G_S}.  Denote by $b_p = \beta_p(G_p)$ the Betti number of the full-depth path homology of $G_p$; we know that $b_p = \dim[\ker(\partial_p)\cap \Cal A_p(G_p)]$.  From
\Cref{t:cycle}, $\ker(\partial_p) \cap \Cal A_p(G_p)\cong \Cal
V_{K_p}$.  This section gives an recursive algorithm to compute a
sequence of matrices $V_{K_p}$, such that the column vectors of each
$V_{K_p}$ form a basis of $\Cal V_{K_p}$.  During the recursion, if
any $\Cal V_{K_p}$ turns out to be $\{0\}$, then by \Cref{p:cycle-sd}
and  induction, $\Cal V_{K_q}=\{0\}$ for all $q\ge
p$, in particular, $\Cal V_{K_L}$ is trivial.  As a result, the
recursion can stop at $p$.  If the recursion reaches $p=L$, then
$\beta_L(G) = \ncol(V_{K_L})$, where $\ncol(M)$ denotes the number
of columns of a matrix $M$.  Denote by $\nrow(\cdot)$ the number of
rows of $M$.

Given $p\ge0$, let $K^+_p$ be as in \eqref{e:S+} and $A_x$ as in
\Cref{t:cycle} for each $x\in K^+_p$.  Let $n_p=|K^+_p|$ and enumerate
$K^+_p$ as $\eno{x^p} {n_p}$.  From \Cref{t:cycle},
\begin{align} \label{e:V-A}
  \def\arraystretch{.7} \arraycolsep=3pt
  \Cal V_{K_p} =
  \Cbr{
    \begin{pmatrix}
      v_1 \\[-.5ex]\vdots \\v_{n_p}
    \end{pmatrix}:
    \sum_{i=1}^{n_p} A_{x^p_i} v_i=0
  }=
  \begin{cases}
    \{0\} & \text{if~} n_p=0 \\
    \Cal N([A_{x_{1}^p}, \ldots, A_{x_{n_p}^p}]) & \text{if~}
    n_p\ge1. 
  \end{cases}
\end{align}
If $\Cal V_{K_p}=\{0\}$, in particular, if
$n_p\le1$, then no further calculation is needed.  Otherwise, a matrix 
\begin{align} \label{e:VKp}
  \def\arraystretch{.7} \arraycolsep=3pt
  V_{K_p} = \begin{pmatrix}
    v_{11} & v_{12} & \ldots & v_{1R} \\
    v_{21} & v_{22} & \ldots & v_{2R} \\
    \vdots & \vdots & \ddots & \vdots \\
    v_{n_p1} & v_{n_p2} & \ldots & v_{n_pR}
  \end{pmatrix}  \quad \text{with~} R = \dim(\Cal V_{K_p})
\end{align}
can be found whose column vectors form a basis of $\Cal V_{K_p}$.
Note that each $v_{ij}$ in $V_{K_p}$ is a vector with dimension equal
to $\ncol(A_{x^p_i})$.

\def\TB{\textsc{TrackBasis}}
\def\Basis{\operatorname{Basis}}
\begin{algorithm}[t!]
  \caption{Computing maximal path homology of a stratified digraph}
  \label{a:recursion}
  \begin{algorithmic}[1]
    \Require Stratified digraph $G$ with layers $\eno[0] K L$, where
    $K_0 = \{\eno{x^0} {n_0}\}$ with $n_0\ge1$; Boolean variable
    $\TB$.
    \Ensure $\beta_L(G)$ and the basis $B\Sp L$ of $H_L(G)$ if
    $\TB=\text{True}$
      \State $b_p\gets0$, $B\Sp p\gets\emptyset$ for all $p=0, \ldots,
      L$.
      \Comment{$b_p = \beta_p(G_p) = \dim[H_p(G_p)]$}
    \If{$n_0=1$}
    \State\Return $b_L$ and $B\Sp L$.
    \EndIf
    \State $A_{x^0_i}\gets1$ for all $i=1, \ldots, n_0$.
    \State  $K^+_0 \gets K_0$, $V_{K_0} \gets [-1_{n_0-1},
    \Id_{n_0-1}]'$, $b_0\gets n_0-1$.
    \If{$\TB$}
    \State $B\Sp0 \gets (x^0_2 -x^0_1, \ldots, x^0_{n_0}-x^0_1)$.
    \EndIf
    \For{$p = 1$ to $L$}
    \State Partition $V_{K_{p-1}}$ as $[\eno{V^{(p-1)\prime}}
    {n_{p-1}}]'$ so that $\nrow(V\Sp {p-1}_i) = \ncol(A_{x^{p-1}_i})$
    for each $i$.  \label{a:partV}
    \State $n_p\gets0$.
    \ForAll{$x\in K_p$}
    \State $I\gets\{i: x^{p-1}_i \in \pred(x)\cap K^+_{p-1}\}$.
    \State $\tilde V_x\gets$ submatrix of $V_{K_{p-1}}$ with all
    $V\Sp{p-1}_i$, $i\in I$, being removed.
    \If{$\Cal N(\tilde V_x)\ne\{0\}$}
    \State $n_p\gets n_p+1$, $x^p_{n_p}\gets x$,  $A_{x^p_{n_p}}\gets
    \Basis(\Cal N(\tilde V_x))$.  \label{a:null-V}
    \EndIf
    \EndFor
    \If{$n_p\le1$ or $\Cal N([A_{x^p_1}, \ldots, A_{x^p_{n_p}}])=\{0\}$}
    \State \textbf{break}
    \EndIf
    \State $K^+_p\gets \{\eno {x^p} {n_p}\}$,  $V_{K_p}\gets
    \Basis(\Cal N([A_{x^p_1}, \ldots, A_{x^p_{n_p}}]))$, $b_p\gets
    \ncol(V_{K_p})$.  \label{a:null-A}
    \If{$\TB$}
    \State $B\Sp p\gets B\Sp{p-1}\sum^{n_p}_{i=1} x_i^p A_{x^p_i} V\Sp
    p_i$.  \label{a:bases}
    \EndIf
    \EndFor
    \State\Return $b_L$ and $B\Sp L$.  \Comment{$B\Sp L=\emptyset$
      if $\TB=$ False}
  \end{algorithmic}
\end{algorithm}

The calculation requires $K^+_p$ and $\{A_x, x\in K^+_p\}$ as input.
These are computed recursively together with $V_{K_p}$.  Start with
$p=0$.  The remark below \Cref {p:cycle-sd} gives $K^+_0=K_0$.  Let
$K_0= \{\eno{x^0} {n_0}\}$.  By \Cref{t:cycle}, set $A_{x^0_i}=1$ for
$i\le n_0$ and provided $n_0>1$, $V_{K_0} = [-1_{n_0-1},
\Id_{n_0-1}]'$, so that by \eqref{e:V-A}, the column vectors of
$V_{K_0}$ form a basis of $\Cal V_{K_0}$, where $1_n$ denotes the
column vector of $n$ 1's and $\Id_n$ the $n\times n$ identity matrix.
To proceed from $p<L$ to $p+1$, let $\Cal V_{K_p}\ne\{0\}$, and assume
that $K^+_p = \{\eno {x^p} {n_p}\}$, $\{A_{x^p_i}, 1\le i\le n_p\}$,
and $V_{K_p}$ are available.  Put $V\Sp p_i = (v_{i1}, \ldots,
v_{iR})$, which  is a submatrix of $V_{K_p}$.  For each $x\in K_{p+1}$,
let $\pred(x)^+ =\pred(x) \cap K^+_p = \{x^p_{i_1}, \ldots,
x^p_{i_k}\}$, where $1\le \seqop i<k\le n_p$.  Let $\imath_x$ be the
natural embedding $\Cal V_{\pred(x)} \hookrightarrow \Cal V_{K_p}$.
Then by \Cref{t:cycle}, $\imath_x\circ \phi_{\pred(x)}$ is a bijection
from $\ker(\partial_p) \cap \Cal A_p(G_{\pred(x)})$ to
\[
  \def\arraystretch{.7} \arraycolsep=3pt
  \tilde{\Cal V}_{\pred(x)} = \Cbr{
    \begin{pmatrix}
      v_1 \\[-.5ex]\vdots \\v_{n_p}
    \end{pmatrix}:
    \sum_{i=1}^{n_p} A_{x^p_i} v_i=0,\
    v_i=0 \text {~for~} i \not\in\{\eno i k\}
  }.
\]
It follows that $x\in K^+_{p+1}\Iff \tilde{\Cal V}_{\pred(x)}\ne\{0\}$.
Furthermore, letting $\tilde V_x$ be the submatrix of $V_{K_p}$ with
$V\Sp p_{i_1}$, \ldots, $V\Sp p_{i_k}$ removed, $\tilde{\Cal
  V}_{\pred(x)} = \{V_{K_p}a: a\in\Cal N(\tilde V_x)\}$ and so $A_x$
can be any matrix whose column vectors form a basis of $\Cal N(\tilde
V_x)$.  Thus $K^+_{p+1}$ and $\{A_x, x\in K^+_{p+1}\}$ are obtained,
so the recursion can continue.  Note that if $\pred(x)^+ = K^+_p$,
then $\tilde V_x$ is empty, so by convention $\Cal N(\tilde V_x) =
\Reals^R$ and $A_x$ can be set equal to $\Id_R$.

If needed, the basis of $\ker(\partial_p)\cap\Cal A(G_p)$ underlying
the recursion can be tracked.  Denote the basis by $B\Sp p$.  By
$K^+_0 = \{\eno{x^0}  {n_0}\}$ and $V_{K_0} = [-1_{n_0-1},
\Id_{n_0-1}]'$, $B\Sp 0 = (x^0_2 - x^0_1, \ldots, x^0_{n_0} - x^0_1)$.
Let $p>0$.  Suppose $B\Sp {p-1} = (\eno{\gamma^{p-1}} {N_{p-1}})$,
$K^+_p = \{\eno {x^p} {n_p}\}$, $\{A_{x^p_i}, 1\le i\le n_p\}$, and
$V_{K_p}$ as in \eqref{e:VKp} are available.  Then by \Cref{t:cycle},
$B\Sp p$ consists of
\[
  \def\arraystretch{.7} \arraycolsep=3pt
  \gamma^p_j=\phi^{-1}_{K_p}\Grp{
    \begin{pmatrix}
      v_{1j} \\ v_{2j}\\\vdots\\ v_{n_pj}
    \end{pmatrix}
  }, \quad j=1,\ldots, R.
\]
For each $j$, $\gamma^p_j = \sum_{x\in K^+_p}\sigma_x x$ with
$\sigma_x \in \Cal C^x_{p-1}$.  If $x = x^p_i$, then from
\Cref{t:cycle}, $\sigma_x = (\eno{\gamma^{p-1}}
{N_{p-1}}) A_{x^p_i}v_{ij} = B\Sp{p-1} A_{x^p_i} v_{ij}$.  Then
$\gamma^p_j = \sum^{n_p}_{i=1} (B\Sp{p-1} A_{x^p_i} v_{ij}) x^p_i =
B\Sp{p-1} \sum^{n_p}_{i=1} x^p_i A_{x^p_i} v_{ij}$.  It follows that
$B\Sp p = (\eno{\gamma^p} R) = B\Sp{p-1} \sum^{n_p}_{i=1}
x^p_i A_{x^p_i} (v_{i1}, \ldots, v_{iR}) =
B\Sp{p-1} \sum^{n_p}_{i=1} x^p_i A_{x^p_i} V\Sp p_i$.

\Cref {a:recursion} is a pseudocode for the above recursive
computation.  In the algorithm, all vectors are treated as column
vectors, and if $\Cal V$ is a vector space in a Euclidean space, then
$\Basis(\Cal V)$ denotes a matrix whose column vectors form a basis of
$\Cal V$. 

The complexity of the algorithm is mainly due to the computation of
$\Basis(\Cal N(\tilde V_x))$ on line \ref{a:null-V} and $\Basis(\Cal
N([A_{x^p_1}, \ldots, A_{x^p_{n_p}}]))$ on line \ref{a:null-A} and, in
the case of basis tracking, the evaluation of the bases on line
\ref{a:bases}.  The other operations in the algorithm are
insignificant comparing to the above computations, because all of them
are assignment operations except for the one on line \ref{a:partV},
which partitions matrix $V_{K_{p-1}}$.  To estimate the total
complexity of the operations on lines \ref{a:null-V} and
\ref{a:null-A}, recall that for a matrix $M$ of $r$ rows and $c$
columns, the computational complexity to find $\Basis(\Cal
N(M))$ using Gaussian elimination is $O(rc\min(r,c))$.  It can be seen
that for  $x\in K_p$, $\nrow(\tilde V_x)
\le \nrow(V_{K_{p-1}})= \sum_{x\in K^+_{p-1}} \ncol(A_x)\le |K_{p-1}|
b_{p-1}$.  On the other hand, $\ncol(\tilde V_x) = \ncol(V_{K_{p-1}})
= b_{p-1}$.  Therefore, the complexity to compute $\Basis(\Cal N(\tilde
V_x))$ for all $x\in K_p$ is $O(|K_p| |K_{p-1}| b^3_{p-1})$.
Similarly, since for each $x\in K^+_p$, $\ncol(A_x)\le \nrow(A_x) =
\ncol(\tilde V_x) = b_{p-1}$, the complexity to compute $\Basis(\Cal
N([A_{x^p_1}, \ldots, A_{x^p_{n_p}}])$ is $O(|K_p| b^3_{p-1})$.  As a
result, the total complexity to compute the bases is $O(\sum^L_{p=1}
|K_{p-1}| |K_p| b^3_{p-1})$.

\subsection{Auxiliary algorithms} \label{ss:auxiliary}
\def\top{\operatorname{top}}
\def\bottom{\operatorname{bottom}}

Let $G=(V,E)$ be a DAG.  It is known that $\ell(G)$ and the weakly
connected components of $G$ can be found in time $O(|V|+|E|)$ \citep
{sedgewick2011algorithms, pacault1974computing}.  Both can be computed
using functions in the Python package \texttt {NetworkX} \citep
{hagberg2008exploring}, the former by \texttt
{dag\_longest\_path\_length} or \texttt {dag\_longest\_path}, the
latter by \texttt {weakly\_connected\_components}.  On the other
hand, we have not found an existing algorithm to compute the subgraph
$G_*$ defined in \Cref{p:subgraph}, which is a stratified digraph with
the same $\ell(G)$-dimensional path homology group as $G$.  However,
following the method to compute $\ell(G)$, in particular, topological
sorting and edge relaxation \citep {sedgewick2011algorithms}, $G_*$
can be found using \Cref{a:auxiliary}, which is based on the following
observation.  For $v\in V$, let $\top[v]$ (resp.\ $\bottom[v]$) be the
maximum length of an allowed elementary path in $G$ that ends at
(resp.\ starts from) $v$.  Then $\top[v]$ and $\bottom[v]$ can be
found respectively by a forward traversal and a backward traversal
according to the topological order of the vertices, and a directed
edge $(u,v)\in E$ is in $G_*$ if and only if $\top[u] + \bottom[v] =
\ell(G)-1$.  Similar to the computation of $\ell(G)$, \Cref
{a:auxiliary} finds $G_*$ in time $O(|V|+|E|)$.

\def\TS{\textsc{TopologicalSort}}
\begin{algorithm}[t!]
  \caption{Computing the stratified subgraph consisting of longest paths
    of a DAG}
  \label{a:auxiliary}
  \begin{algorithmic}[1]
    \Require DAG $G = (V,E)$ and $\ell(G)$.
    \Ensure $G_* = \cup_{\gamma\in \Cal A_{\ell(G)}} \sppt(\gamma)$
    and its layers as in \Cref{p:subgraph}.
    \If{$\ell(G)=0$}
    \State \Return $G_*=(V,\emptyset)$ and $K_0 = V$.
    \EndIf
    \State $\TS\gets$ the topological order of $v\in V$
    \ForAll{$v\in V$} 
    \State $\top[v] \gets 0$, $\bottom[v]\gets0$.
    \EndFor
    \ForAll{$v$ in $\TS$} \Comment{Forward traversal}
    \State $\top[v] \gets \max\{\top[v], \top[u]+1, u\in \pred_G(v)\}$.
    \EndFor
    \ForAll{$v$ in reversed($\TS$)} \Comment{Backward traversal}
    \State $\bottom[v] \gets \max\{\bottom[v], \bottom[u]+1, u\in
    \succ_G(v)\}$.
    \EndFor
    \State $E_* \gets \{(u,v)\in E: \top[u] + 1 + \bottom[v]=
    \ell(G)\}$.
    \State $V_*\gets\{u\in V: u \text{~is a vertex of an edge in~}
    E_*\}$.
    \For{$i=0$ to $\ell(G)$}
    \State $K_i \gets \{u\in V_*: \top[u] = i\}$.
    \EndFor
    \State \Return $G_*=(V_*, E_*)$ and $\eno[0] K {\ell(G)}$.
  \end{algorithmic}
\end{algorithm}

\section{Experiments}\label{s:experiments}
\subsection{Technical setup}\label{s:experiment-setup}
This section reports numerical experiments that compared \Cref
{a:recursion} and the general algorithm of \cite{chowdhury2021path} and
numerical experiments that computed Betti curves of full-depth path homologies of nested stratified
digraphs.  The algorithms were implemented in Python.  Our script for
\Cref{a:recursion} can be accessed at \url {https://github.com/zhengtongzhu/DAG\_MaxPathHomology}.  For the
general algorithm, the script furnished by \cite{pyproject2022} was
used.

To reduce computation and compare the two
algorithms on an equal footing, given a DAG $G$, \Cref{a:auxiliary}
was first applied to reduce it into a stratified digraph $G_*$, which
was then pruned using \Cref{a:prune} in \Cref{s:pruning}.  Afterwards,
the weakly connected components of the pruned digraph were found.
\Cref{a:recursion} and the general algorithm were then applied to
calculate the full-depth cycle spaces of the components, whose direct
sum is equal to the maximal path homology of $G$. 

Our implementation of \Cref{a:recursion} used sparse representations of the numerical matrices arising in the recursive computation, which included $A_{x^p_i}$, $V_{K_p}$, their submatrices, and matrices obtained from them by column or row concatenation. The compressed sparse row format provided by the open source SciPy library was used as the primary storage format.  Row restrictions, sparse identity matrices, and horizontal and vertical block concatenations were implemented using the sparse-matrix operations of SciPy without forming dense copies of the corresponding matrices.  Coordinate-format representations were used temporarily only when sparse outputs were assembled or their nonzero entries were iterated over.

Null-space bases of matrices such as those on lines \ref{a:null-V} and \ref{a:null-A} in \Cref{a:recursion} were computed using a sparse integer-elimination procedure.  Each row of a matrix was represented by a dictionary containing only its nonzero entries, together with a column-to-row incidence structure used to identify candidate pivot rows. Intermediate row operations were performed using Python integers.

In the construction of the symbolic basis-update matrices on line \ref{a:bases} in \Cref{a:recursion}, the numerical block products were computed using SciPy sparse matrices.  The symbolic output was stored as a SymPy sparse matrix and only its nonzero symbolic entries were explicitly retained.

All experiments were performed on a desktop computer running 64-bit Microsoft Windows 11, equipped with an Intel Core i7-9700 processor with 8 physical cores and 8 logical
processors at a base clock frequency of 3.00 GHz, and 16 GB of DDR4 memory.

\subsection{Comparison of runtime and peak traced allocations}

In each experiment, we first specified a stratified digraph with layers $\eno[0] K L$ whose
vertices in adjacent layers were fully connected, i.e., the set of edges is $\{(u,v): u\in K_i, v\in K_{i+1}, 0\le i<L\}$.  The graph will be
referred to as a base graph.  The number of edges in the graph is $\sum^L_{i=1} |K_{i-1}||K_i|$ and the number of allowed elementary paths with maximum length is $|K_0|\cdots |K_L|$.
We then generated subgraphs by randomly deleting a fraction $1-\rho$ of the edges between each pair of adjacent layers $K_i$ and $K_{i+1}$ in the base graph while keeping all its vertices.  For each base graph, $\rho$ took values $0.1k$ with $1\le k\le 9$.  

In the runtime experiments, given a base graph, 200 subgraphs were independently generated for each value of $\rho$.  For each of the subgraphs, its full-depth path homology was computed by \Cref{a:recursion} and the general algorithm.  In runs of \Cref{a:recursion} without basis
tracking, only the Betti number of the full-depth homology was computed.  In runs of \Cref{a:recursion} with basis tracking, explicit path representatives of the basis cycles were additionally constructed.  The total time it took to finish the computation for the 200 subgraphs was recorded for each algorithm.

In addition to runtime, we compared the memory allocation of the runs.  The procedure is similar to the runtime experiments, except that for each value of $\rho$, only 10 subgraphs were independently generated.  Memory allocation was measured using Python's \texttt{tracemalloc} module, and each run was completed in a freshly spawned process.  Package imports and the construction of the input graph were completed before memory tracing began, and the maximum amount of traced memory allocated during the algorithm call was recorded.  The reported value is the mean peak-traced allocation across the 10 randomly generated subgraphs, expressed in MiB.

The values reported by \texttt{tracemalloc} measure memory
allocations tracked by the Python memory-tracing system and should
not be interpreted as the total resident physical memory of the
process.  In particular, memory allocated internally by external
native libraries may not be completely represented.

We carried out experiments using the following base graphs:
\begin{align*}
  &\Gamma_1: L=1, |K_i| =10 \text{~for~} i=0,1,\\
  &\Gamma_2: L=2, |K_i| =10 \text{~for~} i=0,1,2,\\
  &\Gamma_3: L=3, |K_0|=4, |K_i| =10 \text{~for~} i\ge1,\\
  &\Gamma_4: L=4, |K_0|=4, |K_i| = 10 \text{~for~} i\ge1,\\
  &\Gamma_5: L=5, |K_0|=4, |K_i| = 10 \text{~for~} 1\le i\le4, |K_5| = 5,\\
  &\Gamma_6: L=6, |K_0|=4, |K_i| = 10 \text{~for~} 1\le i\le4, |K_i| = 5 \text{~for~} i\ge5,\\
  &\Gamma_7: L=7, |K_0|=4, |K_i| = 10 \text{~for~} 1\le i\le4, |K_i| = 5 \text{~for~} i\ge5, \\
  &\Gamma_8: L=8, |K_0|=4, |K_i| = 10 \text{~for~} 1\le i\le4, |K_i| = 5 \text{~for~} i\ge5,\\
  &\Gamma_9: L=9, |K_0|=4, |K_i| = 10 \text{~for~} 1\le i\le4, |K_i| = 5 \text{~for~} i\ge5, \\
  &\Gamma_{10}: L=10, |K_0|=4, |K_i| = 10 \text{~for~} 1\le i\le4, |K_i| = 5 \text{~for~} i\ge5.
\end{align*}
For the base graphs $\Gamma_1$, $\Gamma_2$, and $\Gamma_3$, we
implemented both the general algorithm and \Cref{a:recursion}.  For the
latter, both the option of no basis tracking and the option of basis
tracking were implemented.  The results are summarized in
\Cref{t:runtimes}.  As expected, as the depth $L$ increased, the total
runtime increased for both the general algorithm and
\Cref{a:recursion}.  However, when bases were not tracked,
\Cref{a:recursion} was consistently faster and its gain in speed was
significant for deeper and denser subgraphs.  For example, for
subgraphs sampled from $\Gamma_3$ at $\rho=0.9$, it took the general
algorithm more than 3 hours to complete the computation, but
\Cref{a:recursion} less than 5 seconds.  In addition, when bases were
not tracked, the total runtime of \Cref{a:recursion} decreased when
$\rho$ was larger.  One reason is that as $\rho$ increases, each
vertex in a sampled subgraph is more likely to be connected to all the
vertices in the previous layer.  For any such a vertex $x$, $A_x$ in
\Cref{t:cycle} is an identity matrix, say $\Id_k$.  If $x\in K_p$,
then from \eqref{e:V-A}, by appropriately indexing the vertices in
$K_p$, $\Cal V_{K_p}$ can be expressed
as $\Cal N([\Id_k, B])$, where the matrix $B$ combines all $A_{x^p_i}$
with $x^p_i\ne x$.  As a result, $V_{K_p}$ can be set equal to
$[-B',\Id_c]'$ with $c=\ncol(B)$.  This analytic solution of $V_{K_p}$
avoids the use of Gaussian elimination to compute the cycle space,
hence significantly reducing the total computation time.

\begin{table}[t!]
  \begin{center}
  \makebox[\textwidth][c]{
    \begin{tabular}{|c||ccc||ccc||ccc|}
      \hline
      \multirow{2}{*}{$\rho$} & \multicolumn{3}{c||}{$\Gamma_1$}
      &\multicolumn{3}{c||}{$\Gamma_2$}&\multicolumn{3}{c|}{$\Gamma_3$}
      \\
      \cline{2-10}
      & G & R$_{\rm no~track}$ & R$_{\rm track}$ 
      & G & R$_{\rm no~track}$ & R$_{\rm track}$ 
      & G & R$_{\rm no~track}$ & R$_{\rm track}$ \\\hline
      0.1&2.5134e-1&1.0106e-1&4.9184e-1 &1.1209e0&1.1234e-1&1.2651e-1 
      &1.9013e0&1.4883e-1&1.4701e-1\\
      0.2&5.2568e-1&4.6742e-1&7.7715e-1 &4.5689e0&2.1917e-1&3.0198e-1 
      &1.2359e1&2.8281e-1&2.0237e-1\\ 
      0.3&8.3188e-1&7.1194e-1&1.2988e0 &1.4865e1&1.2749e0&2.2826e0 
      &4.2270e1&3.0968e-1&3.2847e-1\\ 
      0.4&1.1583e0&7.9444e-1&1.4770e0 &3.8269e1&1.9291e0&4.0842e0 
      &1.6712e2&1.4357e0&2.3047e0\\ 
      0.5&1.4567e0&8.2664e-1&1.6057e0 &8.3582e1&2.1436e0&5.9780e0 
      &4.8424e2&1.9997e0&4.7078e0\\ 
      0.6&1.9254e0&8.8609e-1&1.7347e0 &1.5800e2&2.3140e0&7.6956e0 
      &1.2436e3&2.4702e0&8.1329e0\\ 
      0.7&2.2400e0&8.5014e-1&1.6431e0 &2.5602e2&2.2574e0&8.3419e0 
      &2.8279e3&2.7945e0&1.3142e1\\ 
      0.8&3.2251e0&1.0399e0&2.0792e0 &4.2406e2&1.9065e0&9.9876e0 
      &5.6753e3&2.5518e0&1.7701e1\\ 
      0.9&3.4927e0&7.9182e-1&1.7755e0 &6.5147e2&1.5792e0&1.1066e1 
      &1.1274e4&1.9644e0&2.5072e1\\
      \hline 
    \end{tabular}
    }
  \end{center}
  \caption{A comparison of runtime to compute full-depth path
    homologies.  Given $\rho$, 200 subgraphs were generated from a
    base graph $\Gamma_i$, each by randomly sampling without
    replacement a fraction $\rho$ of the edges between every pair of
    adjacent layers in $\Gamma_i$.  Then the full-depth path
    homologies of the subgraphs were computed using the general
    algorithm (G), the recursive \Cref{a:recursion} without basis
    tracking (R$_{\rm no~track}$), and \Cref{a:recursion} with basis
    tracking (R$_{\rm track}$).  The total computation time of each
    algorithm is displayed in seconds.
  } \label{t:runtimes}
\end{table}

When explicit bases were tracked, the comparison was somewhat different.  For subgraphs sampled from $\Gamma_1$, the recursive implementation generally required more time than the general algorithm, indicating that symbolic basis construction introduces a non-negligible overhead for graphs with fewer layers.  As the depth increased, however, the recursive implementation became progressively more advantageous.  For $\Gamma_3$, even with basis tracking, the recursive implementation was substantially faster than the general algorithm over the entire range of densities.

The cost of basis tracking is a result of the symbolic transformations added to the recursion to construct explicit path representatives.  Unlike numerical computations, these symbolic operations cannot fully leverage existing computational methods. Consequently, the runtime with basis tracking generally increased with graph density and was consistently larger than the runtime without basis tracking, particularly for $\Gamma_2$ and $\Gamma_3$.

The peak traced allocations for $\Gamma_1$, $\Gamma_2$, and
$\Gamma_3$ are summarized in \Cref{t:memory}. For $\Gamma_1$, the mean peak traced allocation of the general algorithm increased from approximately $0.032$ MiB at $\rho=0.1$ to $0.267$ MiB at $\rho=0.9$, whereas the two recursive implementations remained between approximately $21.6$ and $23.5$ MiB.  Thus, for these graphs with fewer layers, the general implementation required substantially less traced memory.

For $\Gamma_2$, the traced allocation of the general algorithm increased much more rapidly, from approximately $0.063$ MiB to $16.1$ MiB, although it remained below those of both recursive implementations throughout the tested range.  The difference became more pronounced for $\Gamma_3$.  In that case, the traced allocation of the general algorithm increased from approximately $0.078$ MiB at $\rho=0.1$ to $206$ MiB at $\rho=0.9$.  It exceeded both recursive implementations between $\rho=0.6$ and $\rho=0.7$.  At $\rho=0.9$, its peak traced allocation was approximately $9.4$ times that of the recursive implementation without basis tracking and approximately $8.4$ times that of the recursive implementation with basis tracking.

In contrast, the traced allocations of the recursive implementations varied relatively little with either depth or density over these experiments: the values remained between approximately $21.6$ and $24.7$ MiB.  Within the metric recorded by \texttt{tracemalloc}, the results show that the general implementation is more memory-efficient for graphs with fewer layers or sparse structures, whereas the recursive implementation scales substantially better as the graph becomes deeper and denser.

\begin{table}[t!]
  \begin{center}
  \makebox[\textwidth][c]{
    \begin{tabular}{|c||ccc||ccc||ccc|}
      \hline    
      \multirow{2}{*}{$\rho$} & \multicolumn{3}{c||}{$\Gamma_1$}
      &\multicolumn{3}{c||}{$\Gamma_2$}&\multicolumn{3}{c|}{$\Gamma_3$}
      \\
      \cline{2-10}
      & G & R$_{\rm no~track}$ & R$_{\rm track}$ 
      & G & R$_{\rm no~track}$ & R$_{\rm track}$ 
      & G & R$_{\rm no~track}$ & R$_{\rm track}$ \\\hline
      0.1&3.2184e-2&2.1615e1&2.1961e1 &6.3442e-2&2.1625e1&2.1623e1 
      &7.8349e-2&2.1628e1&2.1628e1\\
      0.2&4.8813e-2&2.1636e1&2.3401e1 &1.4112e-1&2.1629e1&2.1629e1 
      &2.2788e-1&2.1634e1&2.1634e1\\ 
      0.3&7.0012e-2&2.1640e1&2.3441e1 &3.5296e-1&2.1670e1&2.3459e1 
      &9.5922e-1&2.1652e1&2.2169e1\\ 
      0.4&9.1043e-2&2.1649e1&2.3464e1 &7.6779e-1&2.1698e1&2.3565e1 
      &3.4793e0&2.1681e1&2.3121e1\\ 
      0.5&1.2082e-1&2.1666e1&2.3484e1 &1.5672e0&2.1765e1&2.3792e1
      &9.5491e0&2.1735e1&2.3626e1\\ 
      0.6&1.4773e-1&2.1667e1&2.3493e1 &3.3157e0&2.1842e1&2.3792e1
      &2.1093e1&2.1870e1&2.3832e1\\ 
      0.7&1.8376e-1&2.1665e1&2.3503e1 &5.9866e0&2.1865e1&2.3859e1
      &4.2875e1&2.1995e1&2.4112e1\\ 
      0.8&2.3004e-1&2.1664e1&2.3503e1 &1.0210e1&2.1785e1&2.3947e1 
      &9.9476e1&2.2137e1&2.4391e1\\ 
      0.9&2.6678e-1&2.1667e1&2.3505e1 &1.6147e1&2.1767e1&2.4069e1 
      &2.0607e2&2.1889e1&2.4673e1\\
      \hline 
    \end{tabular}
    }
  \end{center}
  \caption{A comparison of peak traced allocations during the computation of full-depth path homologies.  For each value of $\rho$, 10 independently generated subgraphs were sampled from each base graph $\Gamma_i$.  The same graph realizations were used for the general algorithm (G), \Cref{a:recursion} without basis tracking (R$_{\rm no~track}$), and \Cref{a:recursion} with basis tracking (R$_{\rm track}$). Each entry is the mean peak traced allocation over the 10 realizations, expressed in MiB.} \label{t:memory}
\end{table}

For the larger base graphs $\Gamma_4$ through $\Gamma_{10}$, the growth observed for $\Gamma_3$ made the general algorithm computationally impractical under the available resources.  We therefore restricted these experiments to \Cref{a:recursion} without basis tracking.  The results are reported in \Cref{t:runtimes2}.  As in
\Cref{t:runtimes}, each entry is the total runtime required to process
200 independently generated subgraphs.

For densities up to $\rho=0.5$, all tested graph families were processed quickly: the total runtime for 200 subgraphs remained below four seconds even for $\Gamma_{10}$.  At $\rho=0.6$, the runtime increased but remained between approximately $5.9$ and $33.5$ seconds.  For $\rho\geq0.7$, however, the effect of depth became substantial.  For $\Gamma_{10}$, the total runtimes were approximately $537$ seconds at $\rho=0.7$, $3.90\times10^3$ seconds at $\rho=0.8$, and $9.92\times10^3$ seconds at $\rho=0.9$, the last of which is approximately $2.8$ hours.

The runtimes for $\Gamma_4$, $\Gamma_5$, and $\Gamma_6$ decreased from $\rho=0.8$ to $\rho=0.9$.  At high densities, vertices are more likely to be adjacent to all vertices in the preceding layer, allowing null spaces to be constructed directly without general sparse elimination.  For $\Gamma_7$ to $\Gamma_{10}$, however, the increasing dimensions of the cycle spaces in the intermediate layer eventually dominated this shortcut, leading to rapid runtime growth at high densities.

\begin{table}[t!]
  \begin{center}
  \makebox[\textwidth][c]{
    \begin{tabular}{|c||c||c||c||c||c||c||c|}
      \hline
      \multirow{2}{*}{$\rho$} & \multicolumn{1}{c||}{$\Gamma_4$}
      & \multicolumn{1}{c||}{$\Gamma_5$}
      & \multicolumn{1}{c||}{$\Gamma_6$}
      & \multicolumn{1}{c||}{$\Gamma_7$}
      & \multicolumn{1}{c||}{$\Gamma_8$}
      & \multicolumn{1}{c||}{$\Gamma_9$}
      &\multicolumn{1}{c|}{$\Gamma_{10}$}
      \\
      \cline{2-8}
      & R$_{\rm no~track}$ & R$_{\rm no~track}$& R$_{\rm no~track}$
      & R$_{\rm no~track}$& R$_{\rm no~track}$& R$_{\rm no~track}$& R$_{\rm no~track}$\\ 
      \hline
      0.1&4.69e-1 &1.86e-1 &1.83e-1 &1.66e-1 &1.63e-1 &2.36e-1 &2.06e-1\\
      0.2&1.81e-1 &2.08e-1 &1.96e-1 &2.27e-1 &2.36e-1 &3.49e-1 &2.98e-1\\ 
      0.3&2.65e-1 &2.77e-1 &2.62e-1 &3.53e-1 &3.03e-1 &4.81e-1 &4.13e-1\\ 
      0.4&1.43e0 &1.34e0 &6.26e-1 &4.46e-1 &4.05e-1 &5.58e-1 &4.81e-1\\ 
      0.5&3.15e0 &3.76e0 &3.46e0 &2.30e0 &1.15e0 &9.11e-1 &7.31e-1\\ 
      0.6&5.88e0 &1.30e1 &1.81e1 &2.33e1 &2.59e1 &3.35e1 &3.13e1\\ 
      0.7&9.34e0 &3.67e1 &7.06e1 &1.24e2 &2.00e2 &4.02e2 &5.37e2\\ 
      0.8&8.97e0 &4.36e1 &1.11e2 &2.33e2 &6.41e2 &1.63e3 &3.90e3\\ 
      0.9&5.46e0 &2.41e1 &8.72e1 &2.61e2 &1.21e3 &3.58e3 &9.92e3\\
      \hline 
    \end{tabular}
    }
  \end{center}
  \caption{Total runtime of \Cref{a:recursion} without basis tracking for the
  larger base graphs $\Gamma_4$ through $\Gamma_{10}$.  For each value
  of $\rho$, 200 independently generated subgraphs were sampled from
  each base graph.  Each entry is the total time, in
  seconds, required to compute the full-depth path homologies of all
  200 subgraphs.} \label{t:runtimes2}
\end{table}

\subsection{Betti curve of full-depth path homology}
\label{ss:nested}
To further test the recursive algorithm, we applied it to compute
Betti numbers of full-depth path homologies of nested stratified
digraphs.  A nested set $\{G_t\}$ of stratified digraphs can be either
decreasing, i.e., $G_s\supset G_t$ for $s<t$, or increasing, i.e.,
$G_s\subset G_t$ for $s<t$.  The plot of the Betti number of the path
homology of $G_t$ as a function of $t$, or Betti curve, is a useful graphical tool in the TDA of digraphs  \citep{chowdhury2019path}.

For Betti numbers \emph{per se\/}, there is no essential difference
between using a decreasing set of digraphs and using an increasing
one, as the two types can turn into each other by reversing the
direction of $t$.  In our experiments, given a base graph $\Gamma$
with $L$ layers, its edges were randomly assigned with weights and
each $G_t$ was constructed as the subgraph of $\Gamma$ that kept
all its vertices but only edges with weights at least $t$.  Then
$\{G_t\}$ is a decreasing set of stratified digraphs each with $L$
layers.  As $G_t$ can change only when $t$ is the weight of an edge,
to compute the Betti curve for the full-depth path homology, we 
only had to compute $\beta_L(G_{t\Sb i})$, where $t\Sb 1 < t\Sb 2 <
\ldots$ are all the different edge weights in increasing order.

\begin{figure}[t!]
  \centering
  \begin{subfigure}[t]{0.45\textwidth}
    \centering
    \includegraphics[width=\textwidth]
    {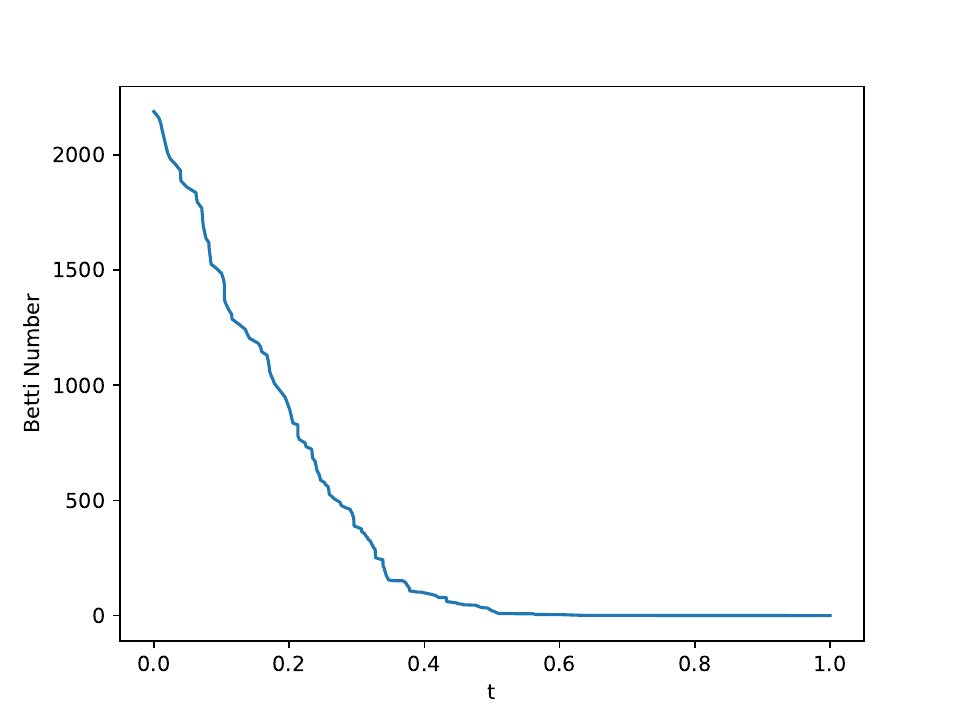}
    \caption{}
    \label{subfig_sec6.2a}
  \end{subfigure}
  \quad
  \begin{subfigure}[t]{0.45\textwidth}
    \centering
    \includegraphics[width=\textwidth]
    {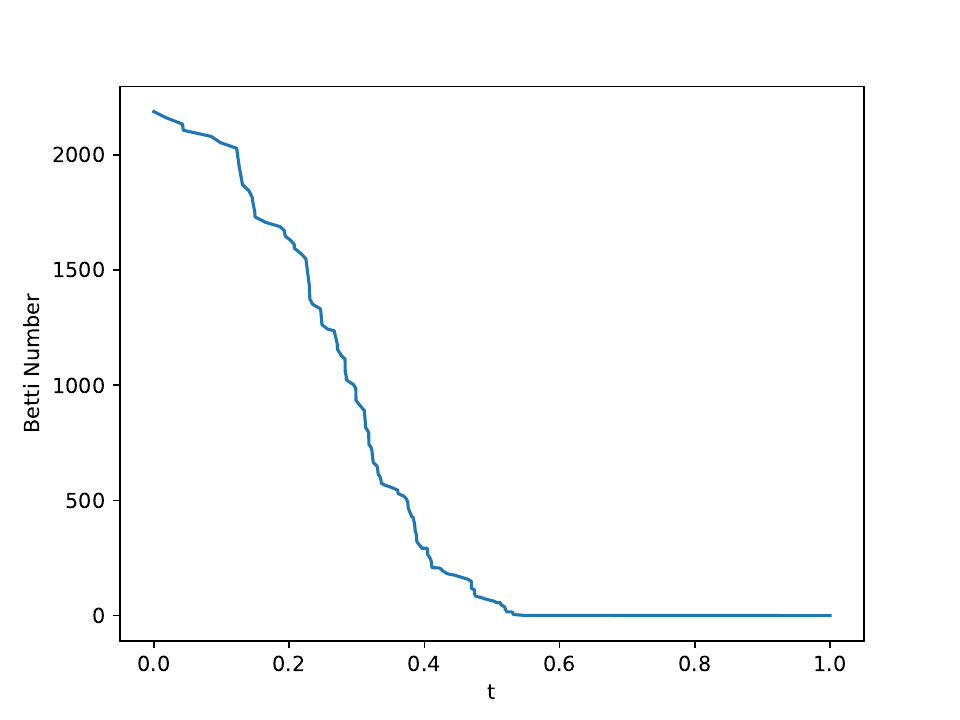}
    \caption{}
    \label{subfig_sec6.2b}
  \end{subfigure}
  \caption{Betti curve for full-depth path homologies.}
  \label{fig_sec6.2}
\end{figure}

The Betti curves $\beta_L(G_t)$ of two sets of nested graphs are
displayed in \Cref{fig_sec6.2}.  In each set, the base graph was
$\Gamma_3$.  For \Cref {subfig_sec6.2a}, the weights assigned to the
edges of $\Gamma_3$ were sampled uniformly from $(0,1)$, while for
\Cref{subfig_sec6.2b}, they were sampled from the Beta distribution
with density $6x(1-x)$, $0<x<1$.  Thus, the range of $t$ in each plot
was set to be $(0,1)$.  In the plots, the Betti curve exhibits a clear
downward, as first pointed out by \cite {chowdhury2019path}.  Indeed,
for $s<t$, by $G_s\supset G_t$ and the fact that each $G_t$ has
depth $L$,
\begin{align} \label{e:nested}
  H_L(G_s) = \ker(\partial_L) \cap \Cal A_L(G_s)
  \supset\ker(\partial_L) \cap \Cal A_L(G_t) = H_L(G_t),
\end{align}
yielding $\beta_L(G_s)\ge \beta_L(G_t)$.  Therefore, the decreasing
monotonicity of $\beta_L(G_t)$ holds in general.

\begin{figure}[t!]
  \centering
  \begin{subfigure}[t]{.27\textwidth}
    \centering
    \setlength{\unitlength}{1mm}
    \begin{picture}(50,45)(-2,0)
      \thicklines
      \put(-2,5){\line(0,1){38}}
      \put(-2,5){\line(1,0){45}}
      \put(-2,2){\tiny0}\put(-2,5){\line(0,-1){1}}
      \put(40,2){\tiny1}\put(40.5,5){\line(0,-1){1}}
      \put(20,2){\tiny$t$}
      
      \put(0,40){\circle*{1}}\put(0,40){\line(1,0){40}}
      \put(-2,40){\line(-1,0){1}}\put(-5,39.5){\tiny1}

      \put(1,37){\circle*{1}}\put(1,37){\line(1,0){39}}
      \put(-2,37){\line(-1,0){1}}\put(-5,36.5){\tiny2}

      \put(4,34){\circle*{1}}\put(4,34){\line(1,0){36}}
      \put(-2,34){\line(-1,0){1}}\put(-5,33.5){\tiny3}

      \put(6,31){\circle*{1}}\put(6,31){\line(1,0){34}}
      \put(-2,31){\line(-1,0){1}}\put(-5,30.5){\tiny4}

      \put(-4,24.2){$\vdots$}\put(15,24){$\ddots$}
      \put(20.7,22.4){$\cdot$}
      \put(-4,19.7){$\vdots$}\put(22,19.7){$\ddots$}

      \put(31,17){\circle*{1}}\put(31,17){\line(1,0){9}}
      \put(-2,17){\line(-1,0){1}}\put(-9.7,16.5){\tiny$n-3$}

      \put(35,14){\circle*{1}}\put(35,14){\line(1,0){5}}
      \put(-2,14){\line(-1,0){1}}\put(-9.7,13.5){\tiny$n-2$}

      \put(37,11){\circle*{1}}\put(37,11){\line(1,0){3}}
      \put(-2,11){\line(-1,0){1}}\put(-9.7,10.5){\tiny$n-1$}

      \put(39,8){\circle*{1}}\put(39,8){\line(1,0){1}}
      \put(-2,8){\line(-1,0){1}}\put(-5,7.5){\tiny$n$}
    \end{picture}
    \caption{Barcode plot for $\tilde G_t$}
    \label{f:barcode}
  \end{subfigure}
  \hspace{1cm}
  \begin{subfigure}[t]{.25\textwidth}
    \centering
    \setlength{\unitlength}{1mm}
    \begin{picture}(50,45)(-2,0)
      \thicklines
      \put(-2,5){\line(0,1){38}}
      \put(-2,5){\line(1,0){45}}
      \put(-2,2){\tiny0}\put(-2,5){\line(0,-1){1}}
      \put(40,2){\tiny1}\put(40.5,5){\line(0,-1){1}}
      \put(20,2){\tiny$t$}
      
      \put(0,40){\circle*{1}}
      \put(-2,40){\line(-1,0){1}}\put(-5,39.5){\tiny1}

      \put(1,37){\circle*{1}}
      \put(-2,37){\line(-1,0){1}}\put(-5,36.5){\tiny2}

      \put(4,34){\circle*{1}}
      \put(-2,34){\line(-1,0){1}}\put(-5,33.5){\tiny3}

      \put(6,31){\circle*{1}}
      \put(-2,31){\line(-1,0){1}}\put(-5,30.5){\tiny4}

      \put(-4,24.2){$\vdots$}\put(15,24){$\ddots$}
      \put(20.7,22.4){$\cdot$}
      \put(-4,19.7){$\vdots$}\put(22,19.7){$\ddots$}

      \put(31,17){\circle*{1}}
      \put(-2,17){\line(-1,0){1}}\put(-9.7,16.5){\tiny$n-3$}

      \put(35,14){\circle*{1}}
      \put(-2,14){\line(-1,0){1}}\put(-9.7,13.5){\tiny$n-2$}

      \put(37,11){\circle*{1}}
      \put(-2,11){\line(-1,0){1}}\put(-9.7,10.5){\tiny$n-1$}

      \put(39,8){\circle*{1}}
      \put(-2,8){\line(-1,0){1}}\put(-5,7.5){\tiny$n$}
    \end{picture}
    \caption{$\beta_L(\tilde G_t)$, upside down}
    \label{f:barcode2}
  \end{subfigure}
  \caption{Barcode plot \emph{vs.\/} Betti curve}
\end{figure}

For full-depth path homology, \eqref{e:nested} also implies a
connection between Betti curve and barcode plot
\citep{zomorodian2004computing, wasserman2018topological, watanabe2022topological}.  To see the
connection, for $0\le t\le1$, let $\tilde G_t = G_{1-t}$.  Then
$\{\tilde G_t\}$ is an increasing set.  From \eqref{e:nested},
$H_L(\tilde G_s) \subset H_L(\tilde G_t)$, $s<t$, which implies that
once a basis element in $H_L(\tilde G_t)$ is born, it lasts forever.
A barcode diagram of the elements is thus as in \Cref{f:barcode},
which consists of horizontal rays each extending from the birth time
of an element to $\infty$.  As usual, elements that are born earlier
have the corresponding rays placed closer to the top.  They are also
indexed in the order of birth time.  Because at the time the $k$th
basis element is born, the Betti number of the full-depth path
homology becomes $k$, the index can also be read as the Betti number.
Now, if we only keep the starting points of the rays in the plot,
then no information is lost.  The resulting plot, as shown
in \Cref{f:barcode2}, is the Betti curve of $\tilde G_t$ at birth
times that is put upside down.

\section{Discussions}\label{s:conclusion}
In this paper, we develop a recursive algorithm to compute the
full-depth path homology of a stratified digraph by taking advantage
of a certain Markovian structure of the homology.  It allows the homology to be computed from numerical matrices without symbolic operations.  As a result, the Betti number of the homology can be computed without the overhead of basis tracking, which is a substantial saving especially for high-dimensional homologies.  On the other hand, the basis elements of the full-depth path homology are encoded in the numerical matrices that are computed recursively, so if needed, they can be tracked by additional symbolic operations during the recursion.  Our numerical experiments showed that with basis tracking, the computational cost of the recursive algorithm was substantially higher, but still substantially lower than the general algorithm.

Unfortunately, for lower-dimensional path homologies, the Markovian structure used in the recursive algorithm breaks down.  To compute lower-dimensional path homologies efficiently, it is necessary to carefully explore their structures, some of which may turn out to be Markovian.  Results such as \Cref{p:decomp3} might provide some useful clues. 

In \Cref{ss:nested}, we report numerical experiments in which the
recursive algorithm was used to compute the Betti curve of
full-depth path homology of a nested set of stratified digraphs.
We also point out that for full-depth path homology, Betti curve and
barcode plot can be derived from each other and basically have the
same visual effect.  Because the latter is regularly used in
persistent homology, the experiments show usefulness of the algorithm
in the area.  On the other hand, a core issue in persistent homology
is to study the births and deaths of basis elements over the course of
a nested set of topological objects, which requires knowing not only
the times when births or deaths occur, but also the basis elements
that are born or die at those times \citep[cf.][]
{edelsbrunner2002topological, masulli2016topology,
  chowdhury2018functorial, wasserman2018topological,
  watanabe2022topological}.  Since the recursive algorithm can only be 
applied  to a fixed digraph at a time, it cannot tell which basis
elements in the homology are born or die when the Betti number
changes.  It is useful to extend the algorithm to address this issue.

We next briefly discuss a few theoretical aspects of
path homology, which may provide useful perspectives to the
algorithm.

First, generalization to non-stratified digraphs.  In \Cref{s:intro},
it was noted that the recursive algorithm can be used to compute the
maximal path homology of a directed acyclic graph (DAG).  Combining
\Cref{p:subgraph} and the results in \Cref {s:algorithms}, one way to
do this is the following.  First, given a DAG $G$, apply \Cref
{a:auxiliary} in \Cref{ss:auxiliary} to find a stratified subgraph
$G_*$ of $G$ whose full-depth path homology is equal to the maximal
path homology of $G$.  Second, apply \Cref {a:recursion} to $G_*$.
The output is then the maximal path homology of $G$.

On the other hand, if a digraph contains directed circles, then it
does not have maximal path homology, as its directed paths can be
arbitrarily long.  Nevertheless, in some cases, high-dimensional path 
homologies of the digraph are still tractable
\citep{grigoryan2012homologies}.  It is therefore of interest to see if
certain high-dimensional path homologies of digraphs with directed
circles can be computed recursively.

Second, stability of path homology.  Generally speaking, full-depth
path homology is not stable under graph perturbations.  For example,
let $G$ be a fully connected stratified digraph with layers $\eno[0] K
L$.  From \cite{chowdhury2019path}, $\beta_L(G) = (|K_0|-1) (|K_1|-1) 
\cdots (|K_L|-1)$.  Suppose $L\ge2$ and $|K_L|=1$.  Then
$\beta_L(G)=0$.  Now add one vertex to $K_L$ and connect it to every
vertex in $K_{L-1}$.  The resulting graph has the $L$th Betti number
equal to $(|K_0|-1)(|K_1|-1) \cdots (|K_{L-1}|-1)$.  If $|K_{L-1}|=2$,
then only one vertex and two edges are added, however, the increment
in the $L$th Betti number, which equals $(|K_0|-1) (|K_1|-1)\cdots  (|K_{L-2}|-1)$, can be arbitrarily large.

On the other hand, in the numerical experiments reported in \Cref
{ss:nested}, the edges in a fully connected stratified digraph were
assigned with random weights and then deleted sequentially according
to the weights.  Thus, the edges were essentially randomly deleted.
In contrast to the above example, from the plots in \Cref{fig_sec6.2},
no sudden changes in the Betti number occurred.  This raises the
question whether the example is an extreme case and the Betti number
of full-depth path homology is actually stable with high probability
under a random perturbation to the graph.  It would be of interest to
address the question in a rigorous way, which could shed more light on
path homologies of digraphs.

Finally, besides path homologies, several other homologies that take into edge orientations have been proposed, for example, directed flag homology \citep{reimann2017cliques} and Dowker homology \citep{chowdhury2018functorial}.  From \cite{chowdhury2019path}, all the second- and higher-dimensional directed flag homologies are trivial for stratified digraphs.  On the other hand, in Dowker homology, vertices can form a simplex only if they have a common predecessor or successor.  Therefore, in a stratified digraph, only vertices belonging to the same layer can form simplies, and so the Dowker homology of the graph is a direct sum of the simplicial homologies of individual layers each regarded as a finite set of points.  Both homologies thus encode less information about the edges of a stratified diagraph than the path homology.  On the other hand, for digraphs with richer structures, both the directed flag homology and Dowker homology may carry more information and it would be interesting to quantify the differences between them and path homology.

\bibliography{references}

\appendix
\section{An algorithm to prune a digraph} \label{s:pruning}
\begin{algorithm}[t!]
  \caption{Pruning of a stratified digraph}
  \label{a:prune}
  \begin{algorithmic}[1]
    \Require Stratified digraph $G_*=(V_*,E_*)$
    \State $G\gets G_*$
    \State $Q\gets$ list of $v\in V_*$ that are removable from $G$
    \While{$Q\ne\emptyset$}
      \State $v\gets$ the first element in $Q$
      \State $S \gets [\pred_G(v)\cup \succ_G(v)]\setminus Q$
      \State update $G$ by deleting $v$ and all the edges to or from
      $v$
      \State remove $v$ from $Q$
      \State add $x\in S$ that are removable from $G$ to the end of
      $Q$
    \EndWhile
    \State\Return $G$
  \end{algorithmic}
\end{algorithm}

\Cref{a:prune} carries out the pruning procedure described after
\Cref{c:decomp}.  Recall that a vertex $x$ in
a stratified digraph $G$ is removable if (1) $x$ is not in the top
layer of $G$ and $|\pred_G(x)|\le1$ or (2) $x$ is not in the bottom
layer of $G$ and $|\succ_G(x)|\le1$.  The algorithm records removable
vertices in a list $Q$, which is used as a queue.  In each iteration,
the first vertex in $Q$ is deleted from $G$ together with its incoming
and outgoing edges.  If there are vertices that become removable due
to the deletion, they are added to $Q$.  Afterwards, the first vertex
is removed from $Q$.  The iteration continues until $Q$ is empty.
Since the algorithm only deletes removable vertices, the output
digraph $G$ has the same space of full-depth cycles as the input
digraph $G_*$.  Assume that $G$ still has a removable vertex $x$.
Then $x$ is not in the initial $Q$ but becomes removable due to the
deletion of some vertex $v$ from $G$.  This can happen only if $x\in
[\pred_G(v) \cup \succ_G(v)]\setminus Q$, which implies that $x$ is
added to $Q$ and eventually deleted.  The contradiction implies that
$G$ has no removable vertices. 

To bound the computational complexity of the algorithm, note that all
the elements in $Q$ are distinct.  If a vertex is in $Q$, it
will stay there until being deleted from $G$ and $Q$.  In 
other words, a vertex can be in $Q$ at most once.  As a result, the
number of iterations of the while loop is at most $|V_*|$.   Within each
iteration, if $v$ is the first vertex in $Q$, then
vertices in $\pred_G(v)\cup\succ_G(v)$ are checked.  As a result,
the total number of times vertices are checked is bounded by
$|V_*|(1 + \sum_{v\in V_*} |\pred_{G_*}(v)| + |\pred_{G_*}(v)|) = 
|V_*| + 2|E_*|$, so the total running time is $O(|V_*|+|E_*|)$.

The memory space used by $Q$ and $S$ in the algorithm is $O(|V_*|)$.  
Including the copied graph $G$, the total complexity of the space is $O(|V_*|+|E_*|)$.
\end{document}